\documentclass{aastex631}
\usepackage{graphicx}
\usepackage{amsmath}
\usepackage{amsthm}
\usepackage{booktabs}
\usepackage{psfrag}
\usepackage{epstopdf}
\usepackage{placeins}

\def\eqp#1{(\ref{eq:#1})}
\def\eql#1{\label{eq:#1}}
\newcommand{\be}{\begin{equation}}
\newcommand{\ee}{\end{equation}}
\newcommand{\ba}{\begin{eqnarray}}
\newcommand{\ea}{\end{eqnarray}}
\newtheorem{proposition}{Proposition}
\newtheorem*{remark}{Remark}
\def\order#1{\mathcal{O}\left(#1\right)}
\newcommand\Omegab{\overline{\Omega}}

\newcommand\phip{\phi'}
\newcommand\rp{r'}
\newcommand\tp{t'}
\newcommand\p{\partial}
\newcommand\ddphi{\frac{\partial}{\partial\phi}}
\newcommand\ddphip{\frac{\partial}{\partial\phip}}
\newcommand\ddt{\frac{\p}{\p t}}

\newcommand\ddtp{\frac{\partial}{\partial\tp}}
\newcommand\ddr{\frac{\partial}{\partial r}}
\newcommand\ddrp{\frac{\partial}{\partial\rp}}

\newcommand\filter{\mathrm{filter}}

\newcommand\epsfilter{\epsilon_\filter}
\newcommand\bfu{\mathbf{u}}

\newcommand\vecu{\vec{u}}
\newcommand\mypart{p_\mathrm{art}}
\newcommand\myCap{C_\mathrm{ap}}
\newcommand\rmL{\mathrm{L}}
\newcommand\rmR{\mathrm{R}}
\newcommand\rhoL{\rho_\rmL}
\newcommand\uL{u_\rmL}
\newcommand\pL{p_\rmL}
\newcommand\rhoR{\rho_\rmR}
\newcommand\uR{u_\rmR}
\newcommand\pR{p_\rmR}
\newcommand\CFL{\mathrm{CFL}} 

\newcommand\rmv{\mathrm{v}}
\newcommand\rmi{\mathrm{i}}
\newcommand\eint{e_\mathrm{int}}

\newcommand\half{\frac{1}{2}}
\newcommand\eqspace{\hskip 0.25truecm}
\newcommand\ri{\frac{1}{r}}

\newcommand\Rey{\mathrm{Re}}
\newcommand\Rei{\Rey_\rmi}
\newcommand\Reo{\Rey_\rmo}
\newcommand\rmo{\mathrm{o}}
\newcommand\rinner{r_\rmi}
\newcommand\router{r_\rmo}
\newcommand\Omegai{\Omega_\rmi}
\newcommand\Omegao{\Omega_\rmo}
\newcommand\rmb{\mathrm{b}}

\newcommand\pt{\frac{\p}{\p t}}
\newcommand\pz{\frac{\p}{\p z}}
\newcommand\pr{\frac{\p}{\p r}}
\newcommand\pphi{\frac{\p}{\p\phi}}
\newcommand\bfb{\mathbf{b}}
\newcommand\bff{\mathbf{f}}
\newcommand\bfU{\mathbf{U}}
\newcommand\bfw{\mathbf{w}}
\newcommand\bffp{\bff^\prime}

\newcommand\bfwt{\widetilde{\bfw}}
\newcommand\wh{\widehat{w}}
\newcommand\myspace{\hskip 0.25truecm} 
\newcommand\rmeff{\mathrm{eff}}
\newcommand\fpr{f^\prime}
\newcommand\what{\widehat{w}}

\newcommand\tauth{t_\mathrm{thermal}}
\newcommand\rmmin{\mathrm{min}}
\newcommand\rmmax{\mathrm{max}}

\newcommand\rmin{r_\rmmin}
\newcommand\rmax{r_\rmmax}
\newcommand\zmin{z_\rmmin}
\newcommand\zmax{z_\rmmax}
\newcommand\ahat{\widehat{a}}
\newcommand\epsfil{\epsilon_\mathrm{filter}}
\newcommand\qveccond{\vec{q}_\mathrm{cond}}
\newcommand\qvec{\vec{q}}

\newcommand\Ngrid{N_\mathrm{grid}}
\newcommand\Ncores{N_\mathrm{cores}}
\newcommand\tcpu{t_\mathrm{cpu}}

\submitjournal{ApJ Supp.}

\shorttitle{Pad\'e code}
\shortauthors{Shariff}

\begin{document}

\title{{\sc Pad\'e}: A code for protoplanetary disk turbulence based on Pad\'e differencing}

\correspondingauthor{Karim Shariff}
\email{karim.shariff@nasa.gov,}

\author[0000-0002-7256-2497]{Karim Shariff}
\affil{NASA Ames Research Center \\
Moffett Field, CA 94035, USA}

\begin{abstract}

The {\sc Pad\'e} code has been developed to treat hydrodynamic turbulence in protoplanetary disks.  It solves the compressible equations of motion in cylindrical coordinates.  Derivatives are computed using non-diffusive and conservative fourth-order Pad\'e differencing, which has higher resolving power compared to both dissipative shock-capturing schemes used in most astrophysics codes, as well as non-diffusive central finite-difference schemes of the same order.  The fourth-order Runge-Kutta method is used for time stepping.  A previously reported error-corrected Fargo approach is used to reduce the time step constraint imposed by rapid Keplerian advection.  Artificial bulk viscosity is used when shock-capturing is required.  Tests for correctness and scaling with respect to the number of processors are presented.  Finally, efforts to improve efficiency and accuracy are suggested. 

\end{abstract}


\section{Introduction}

This paper presents a code called {\sc Pad\'e}, developed to simulate hydrodynamic turbulence in protoplanetary disks.  The code uses a Pad\'e/compact finite-difference scheme \citep{Lele_1992}.  Such schemes have spectral-like resolving power.  This means that they approach the ability of a spectral method to compute derivatives exactly for all wave numbers that the mesh can support.  In particular, for advection problems they have zero diffusive error (as do all central finite-difference schemes) and have small dispersion error across the wavenumber range.  Thus they have the ability to more accurately treat the dynamics of the small scales supported by the mesh and produce energy spectra that have a wider inertial (power-law) range for turbulent flows.

On the other hand, most astrophysical codes employ Godunov-type schemes that were an elegant and mathematically supported breakthrough for capturing shock-waves.  Such schemes employ non-central upwinded finite-difference or flux reconstruction schemes with a smart non-linear dissipation that is necessary for capturing shocks, but which leads to excessive dissipation of vortical and other smooth features such as density waves.
To fix this issue there have been attempts along many directions which include hybrid methods \citep{Adams_and_Shariff_1996, Pirozzoli_2002}, non-linear filtering \citep{Yee_and_Sjogreen_2018}, and vorticity-preserving schemes \citep{Lerat_etal_2007, Seligman_and_Laughlin_2017, Seligman_and_Shariff_2019}.  
Since shocks have not been observed in protoplanetary disk turbulence to date, we can side step the issue.
For treating shock waves that are not too strong, the {\sc Pad\'e} code provides an optional artificial bulk viscosity treatment \citep{Cook_and_Cabot_2005, Mani_etal_2009}.  This is not as elegant or oscillation free as Godunov methods, but is designed to apply a diffusivity only where the divergence $\nabla \cdot \bfu$ is strong.
\newpage

The development of the {\sc Pad\'e} code was motivated by the discovery in the last two decades of a number of mechanisms for the generation of hydrodynamic turbulence in protoplanetary disks; see \cite{Lesur_etal_2023} for a comprehensive review.
These include the vertical shear instability (VSI), convective over-stability (COS), and the zombie vortex instability (ZVI). Each is most strongly amplified for a different range of $\Omega\,\tauth$, the time scale for radiative relaxation of temperature fluctuations back to the background (normalized by the orbital frequency, $\Omega$).  VSI is most strongly amplified when $\Omega\,\tauth \ll 1$, COS when $\Omega\,\tauth \sim 1$, and ZVI when  $\Omega\,\tauth\gg 1$.  Turbulence can also be driven by the magneto-rotational instability (MRI) in the radially inner and outer regions of protoplanetary disks where ionization is sufficient.

As mentioned earlier, most astrophysical codes that are applied to protoplanetary disks use dissipative shock-capturing methods.  These codes include {\sc Athena} \citep{Stone_etal_2008}, {\sc Athena++} \citep{Stone_etal_2020}, {\sc Pluto} \citep{Mignone_etal_2007}, and
{\sc Fargo3D} \citep{Benitez_Llambay_etal_2016}.
The highest order scheme provided in {\sc Athena} and {\sc Athena++} is the third order PPM (Piecewise Parabolic Method).
The highest order scheme in {\sc Pluto} is a fifth order WENO (Weighted Essentially Non-Oscillatory) finite difference shock-capturing scheme denoted WENOZ\_FD in the {\sc Pluto} manual \cite[][pg. 93]{Pluto_manual_2021}.  
{\sc Fargo3d} employs a staggered mesh such that the scalar variables, density and internal energy per unit volume, are cell-centered while vector quantities, velocity and magnetic field, are located at the centers of cell faces.  The computation of fluxes at cell faces employs upwinded interpolation which is necessarily dissipative.  Time advancement uses operator splitting whereby different sets of terms contributing to the time rate of change of flow quantities are time-advanced separately, one after the other.

The one exception to the use of dissipative shock-capturing scheme is the {\sc Pencil} code \citep{Brandenburg_etal_2021} which uses sixth-order central differencing.  Central schemes, including the Pad\'e scheme, produce oscillations at the Nyquist wavenumber of the mesh.  To overcome this, {\sc Pencil} uses a dissipative fifth order upwind biased scheme, while {\sc Pad\'e} uses a filter with a sharp cut-off (\S\ref{sec:filter}).

The rest of the paper presents the equations solved (\S\ref{sec:eqs}), computational schemes (\S\ref{sec:numerics}), various tests (\S\ref{sec:tests}), and closing remarks (\S\ref{sec:close}).  Code availability is described after the acknowledgements.


\section{Equations discretized}\label{sec:eqs}

\subsection{Transport equations}

The code solves the equations for mass and momentum (radial, angular, and vertical) transport written in as close to flux-divergence form as possible in cylindrical coordinates
\ba
&&\frac{\p\rho}{\p t} + \pz(\rho u_z) + \frac{1}{r}\pr(r\rho u_r)+ \frac{1}{r} \pphi(\rho u_\phi) = 0, \eql{mass} \\
&&\pt(\rho u_r) + \pz(\rho u_r u_z) + \frac{1}{r}\pr\left[r (p + \rho u_r^2)\right] + \frac{1}{r} \pphi(\rho u_\phi u_r) - \frac{\rho u_\phi^2}{r} =
\frac{p}{r} + \rho g_r + F^\rmv_r, \eql{rmom} \\
&&\pt(\rho u_\phi r) + \pz(\rho u_\phi r u_z) + \frac{1}{r} \pr(r\rho u_\phi r u_r) + \pphi(p + \rho u_\phi^2) = 
r F^\rmv_\phi,\\
&&\pr(\rho u_z) + \pz(p + \rho u_z^2) + \frac{1}{r}\pr(r\rho u_z u_r) + \frac{1}{r}\pphi (\rho u_z u_\phi) = 
\rho g_z +  F^\rmv_z.
\ea
Here $\mathbf{F}^\rmv$ represents an optional viscous force given in Appendix~\ref{sec:viscous_terms}.  Implementation of characteristic boundary conditions requires calculation of the flux Jacobian, and to make the flux look similar to that in Cartesian coordinates, the term $-\partial p/\partial r$ that normally appears on the right-hand-side of the radial momentum equation \eqp{rmom}  was moved into the radial advective flux on the left-hand-side.  This is accomplished by adding the quantity
\be
   \frac{1}{r}\pr\left(r p\right) = \frac{1}{r} \left(r\frac{\p p}{\p r} + p \right) = \frac{\p p}{\p r} + \frac{p}{r},
\ee
to both sides of \eqp{rmom}, which results in the source term $p/r$ on the right-hand-side.

For the locally isothermal option, the pressure is computed as $p = \rho c_\rmi^2(r)$ where $c_\rmi(r, z)$ is the local isothermal sound speed.  For the non-isothermal option we have
\be
   p = (\gamma - 1) \eint,
\ee
where $\eint$ is the internal energy (per unit volume).  The rationale for using the internal energy instead of its sum with the kinetic energy is given in our work on the Fargo method \citep{Shariff_and_Wray_2018}.  The transport equation for $\eint$ is
\be
\frac{\p\eint}{\p t} + \pz(\eint u_z) + \frac{1}{r}\pr(r\eint u_r) + \frac{1}{r}\pphi(\eint u_\phi) = -p\nabla\cdot \vecu + Q^\rmv - \nabla\cdot\qveccond. \eql{eint}
\ee
The first term on the right-hand-side of \eqp{eint} is the pressure-dilatation term which causes heating under compression.  The dilatation is given by
\be
   \nabla\cdot\vecu = \frac{\p u_z}{\p z} + \frac{1}{r}\pr(r u_r) + \frac{1}{r}\frac{\p u_\phi}{\p\phi}
\ee
The last two terms $Q^\rmv -\nabla\cdot\qveccond$ on the right-hand-side of \eqp{eint} represent viscous heating and conductive heat transfer, respectively.  They are activated only if viscous terms are activated; their form is given in Appendix \ref{sec:viscous_terms}.

The variables evolved at each grid point are $\vec{q} = (\rho, \rho u_r, \rho u_\phi r, \rho u_z, \eint)$.
When spatial derivatives are approximated by Pad\'e finite-differences (discussed below), one obtains a system of ODEs for the time rate of change, $\p_t\qvec$ of $\vec{q}$ at each grid point.  This system is evolved using the fourth-order Runge-Kutta method.
Note that we do \textit{not} employ operator splitting whereby different sets of terms in $\p_t\qvec$ are time evolved separately one after another; this would produce a splitting error which we do not incur.

Currently, the main source of gravity in the code is from a mass $M$ at the origin so that
\be
   g_r =-\frac{GM}{R}\frac{r}{R}, \eqspace g_z = -\frac{GM}{R}\frac{z}{R}, \eqspace R \equiv \left(r^2 + z^2\right)^{1/2},  \eql{gravity}
\ee
which are pre-computed.  The code also allows other simple choices such as uniform $g_z$ and the thin disk version of \eqp{gravity}.

\subsection{Fargo}\label{sec:fargo}

The Fargo method was introduced by \cite{Masset_2000} to alleviate the time step restriction resulting from fast Keplerian advection.  \cite{Shariff_and_Wray_2018} improved its accuracy by first noting that underlying the method is a transformation of the azimuthal coordinate $\phi$
\begin{align}
   \phip &= \phi - \Omegab(r)(t - t_0), \\
   \rp &= r, \\
   \tp &= t,
\end{align}
for the duration of a time step, $t_0 < t \leq t_0 + \Delta t$.  Here $\Omegab(r)$ is a prescribed rotation rate that one wishes to subtract from determining the time step.   At the end of the time step, one brings the flow field back to original coordinates by performing a shift using an FFT.    The chain-rule for differentiation then implies that every $t$ and $r$ derivative in the transport equations carries an additional term:
\ba
\ddt      &=& \ddtp - \frac{\p\phip}{\p t}\ddphip = \ddtp - \Omegab(r) \ddphip, \eql{ddt}\\
\ddr      &=&  \ddrp - \frac{\p\phip}{\p r}\ddphip = \ddrp - (t - t_0)\frac{\partial\Omegab}{\partial r}\ddphip,\eql{ddr}\\
\ddphi &=& \ddphip. \eql{ddp}
\ea
The second term in $\p/\p t$ serves to remove $\Omegab(r)$ from the azimuthal advection velocity, therefore, it no longer influences the time step.  To see how this works, consider the transformed mass transport equation \eqp{mass}:
\be
\frac{\p \rho}{\p t'} + \frac{1}{r} \frac{\p}{\p\phi}\left(\rho u'_\phi\right) + \frac{1}{r}\left[\frac{\p}{\p r'} + \chi\right]\left(r\rho u_r\right) + \frac{\p}{\p z}\left(\rho u_z\right) = 0, \eql{transformed}
\ee
where
\be
   u'_\phi = u_\phi - \Omegab(r) r
\ee
is a shifted velocity that results in a less restrictive time step.  In \eqp{transformed}, the symbol $\chi$ denotes the operator
\be
   \chi = -(t - t_0) \frac{\p\Omegab}{\p r} \frac{\p}{\p \phi}.
\ee
It arises from the second term on the right-hand-side of the last member of \eqp{ddr}, and
must be added to every $r$ derivative in the transport equations; the code refers to it as the `extra operator.'  The extra operator $\chi$ is missing in the original implementation \citep{Masset_2000}: its neglect results in an error of $\order{\Delta t}$.  The original implementation also ``integerizes'' the prescribed $\Omegab(r)$ to allow a shift of the flow field by an integer number of grid intervals in $\phi$ at the end of a time step.  The integer shift jumps at certain radial locations and this results in additional error at those locations.

This code provides options to completely or partially revert to Masset's original algorithm if desired.  In particular, integer shifting can be selected as opposed to the more accurate real-valued shifting.  Also, inclusion of the extra operator $\chi$ can be suppressed.  At the end of each time step, the flow field is shifted back to original coordinates.  Real-valued shifting is performed using an FFT from the FFTW library \citep{FFTW05}.

Following publication of \cite{Shariff_and_Wray_2018}, we were contacted by Prof. Pablo Ben\'itez Llambay (Universidad Adolfo Ib\'a\~nez, Chile) who is a main developer of {\sc Fargo3d}.  He pointed out that {\sc Fargo3d} performs advection by operator splitting, and therefore there is no coordinate transformation requiring chain-rule terms.  For instance, consider the continuity equation with only $\phi$ advection for simplicity:
\be
\frac{\p \rho}{\p t} + \frac{1}{r} \frac{\p}{\p\phi}\left(\rho u_\phi\right) = 0,
\ee
with the decomposition
\be
   u_\phi = \Omegab(r) r + u_\phi^\prime.
\ee
In the operator splitting approach, $\rho$ is first partially evolved according to advection by the residual velocity $u_\phi^\prime$,
\be
\frac{1}{r} \frac{\p}{\p\phi}\left(\rho u_\phi^\prime\right),
\ee
whose time step restriction is not severe.  Next, the term $\Omegab(r) \p_\phi\rho$, representing advection by Keplerian flow, is treated in {\sc Fargo3d} by real-valued shifting  (Colin McNally, Queen Mary Univ. of London, Private Communication).  Thus, it is true that no coordinate transformation is implied.  However, operator splitting has $\mathcal{O}\left(\Delta t^2\right)$ error which is of the same order as the error made by dropping the extra chain-rule terms.  Thus, it would appear that for higher order time integration schemes, chain-rule terms are necessary for consistency.

\subsection{Artificial pressure for shock-capturing}\label{sec:ap}

Pad\'e schemes, like spectral schemes, are not designed to capture shocks.  However, one may encounter shocks in protoplanetary disks.  Examples include the infall accretion shock \citep{Neufeld_and_Hollenbach_1994}, and bow shocks due to solid bodies moving supersonically relative to the gas.  For these reasons, the code allows for an optional treatment of shocks using artificial bulk viscosity \citep{Cook_and_Cabot_2005, Mani_etal_2009}.  This method results in an \textit{artificial pressure}, $\mypart$, which is then added to the physical pressure.  The actual calculation of $\mypart$ is
\be
   \mypart= - \beta_\Delta \nabla\cdot\vecu,
\ee
where $\beta_\Delta$ is the artificial bulk viscosity.  Note that artificial pressure is positive in regions of compression (dilatation $\nabla\cdot\vecu < 0$) and negative in regions of expansion.  The artificial bulk viscosity is made sensitive to the dilatation as follows:
\be
   \beta_\Delta = \myCap \rho \ell_\mathrm{grid}^2 |\nabla\cdot\vecu|,
\ee
were $\myCap$ is a user specified coefficient and the grid size squared is
\be
 \ell_\mathrm{grid}^2 = 
 \begin{cases}
 \Delta z^2, & \text{\ in 1D};\\
 r\Delta\phi \Delta r, & \text{\ in the planar case};\\
 (r\Delta\phi\Delta r\Delta z)^{2/3}, & \text{\ in 3D}.
 \end{cases}
\ee
Since the absolute value function is not smooth, the Pad\'e filter (\S\ref{sec:filter}) with $\epsfilter = 0.2$ is applied to $\beta_\Delta$.

The artificial pressure term imposes a time step constraint appropriate for a viscous (second derivative) term.  The time step must satisfy
\be
(\lambda_\mathrm{ap})_\mathrm{max}\Delta t < \mathrm{CFL},
\ee
where $(\lambda_\mathrm{ap})_\mathrm{max}$ is the maximum eigenvalue of the bulk viscosity operator and $\mathrm{CFL}$ is the Courant-Friedrichs-Lewy limit specified by the user.  For the fourth-order Runge-Kutta method, $\mathrm{CFL} < \sqrt{8}$ for stability. 
The maximum eigenvalue is estimated as
\be
   (\lambda_\mathrm{ap})_\mathrm{max} = \left(\frac{\pi^2\beta_\Delta}{\rho \ell_\mathrm{grid}^2}\right)_\mathrm{max},
   \eql{lambda_max_ap}
\ee
where a factor of $\pi$ comes from assuming that each numerical derivative has spectral accuracy.  Recall that $\beta_\Delta \propto \ell_\mathrm{grid}^2$ so the grid size actually drops out in \eqp{lambda_max_ap}.
For the shock-tube and density wave test cases we report below, this eigenvalue was a factor of 1.08 and 1.61 larger, respectively,  than the eigenvalue for the Euler terms.

\section{Numerics}\label{sec:numerics}

\subsection{Pad\'e differentiation}

The motivation for Pad\'e differencing (also referred to as compact differencing) is its high \textit{resolving power} (\S\ref{sec:rp}); see \cite{Lele_1992} for a comprehensive presentation and various extensions.  The idea and initial development of such schemes is due to the Czech born astronomer and numerical analyst Zden\v{e}k Kopal around 1959; see the bibliographical notes (item IX-C) in his book \citep{Kopal_1961}.  The basic idea, developed in Chapter 9 of the book, is to write the exact first derivative operator as an exact function of the central difference operator.  This operator function is first expanded in a truncated Taylor polynomial, which results in a conventional difference scheme.  The key idea, however, is to next obtain a rational polynomial approximation (known as a Pad\'e approximant) to the Taylor series.   It is known that Pad\'e\footnote{Henri Pad\'e was a French mathematician who, for his doctoral thesis, studied (c.~1890) the approximation of functions by rational polynomials, now known as Pad\'e approximants.} approximants to ordinary functions have a greater range of accuracy and radius of convergence than a Taylor series. In the present case, one obtains difference schemes with better resolving power.
The simplest scheme, and the one we use both in the interior of non-periodic domains and for periodic domains is
\be
   \alpha f^\prime_{j-1} + f^\prime_j + \alpha f^\prime_{j+1} = \frac{a}{2h}\left(f_{j+1} - f_{j-1}\right), 
   \eql{std_pade}
\ee
where $h$ is the uniform grid spacing, $\alpha = 1/4$ and $a = 3/2$.  
Equations \eqp{std_pade} constitute a tridiagonal system of equations along each line of data in the mesh.
The system is solved efficiently using the Thomas algorithm, which is simply Gaussian elimination applied to a tridiagonal matrix.
Since Gaussian elimination is recursive, namely, operating on row $j$ depends on the result of row $j-1$, the memory cache cannot be preloaded with the required data.  Similarly, any available vectorization hardware cannot be engaged by the compiler.  To overcome this, we follow the standard practice of having each step of the Thomas algorithm inner-loop over a bundle of independent inversions for different grid lines of the mesh.
Equation \eqp{std_pade} is fourth-order accurate, i.e., its truncation error is $\order{h^4}$.  To allow for non-uniform meshes, numerical differentiation is performed with respect to the continuous grid index variable $\xi$ (such that $\xi_j = j$ and $h = 1$), and the chain rule is used, e.g.,
\be
   \frac{\p f}{\p z} = \frac{\p f}{\p \xi} \left(\frac{dz}{d\xi}\right)^{-1}.
\ee

Kopal's operator calculus is abstruse and unwieldy, but once the basic form of Pad\'e schemes is recognized, an easier approach to develop them is to simply write down a specific form with a desired grid-point stencil, and obtain the unknown coefficients by setting the Taylor series error to zero at various orders.  This is the approach followed in \cite{Lele_1992} who developed a number of extensions, including resolving power optimization, boundary schemes, higher derivatives, conservation, and filtering.

For robustness, it is desirable that a code maintain positivity of density and internal energy.  
Negative values can arise in strongly evacuated regions for low-order schemes, and near very strong discontinuities for higher order schemes;  see, for example, \citet{Hu_etal_2013} who present a simple method for ensuring positivity for finite-difference schemes that can be written as a difference of numerical fluxes.  
The present method does not guarantee positivity, however, the code has not encountered difficulties for problems of subsonic turbulence for which it is intended.  In the future, it may be possible to implement the method of \citet{Hu_etal_2013} using the ``reconstruction by primitive function'' trick  \citep{Harten_etal_1987, Shu_and_Osher_1989}, which is also discussed in \cite{Merriman_2003}.  In this method, one obtains numerical fluxes by differentiating the primitive function; this differentiation would be performed using the Pad\'e scheme.  

\subsection{Resolving power of Pad\'e differencing}\label{sec:rp}

The advantage of Pad\'e schemes is, first of all, their compactness:  for \eqp{std_pade}, for instance, fourth-order accuracy is obtained with a stencil width of three rather than four in the case of standard central differencing.  More importantly, they have better resolving power than standard central differencing.  This means that for the same formal order of accuracy, they provide an accurate derivative up to higher wave numbers.  This is quantified by the so-called \textit{effective wave number} analysis: substitute
\be
   f_j = \exp(\rmi k x_j) \myspace \mathrm{and} \myspace  f^\prime_j = \rmi k_\rmeff(k) \exp(\rmi k x_j), \myspace
   0 \leq k \leq \pi/h,
\ee
into \eqp{std_pade} to obtain the effective wavenumber $k_\rmeff(k)$.
\begin{figure}
\begin{center}
\includegraphics[width=3.5in]{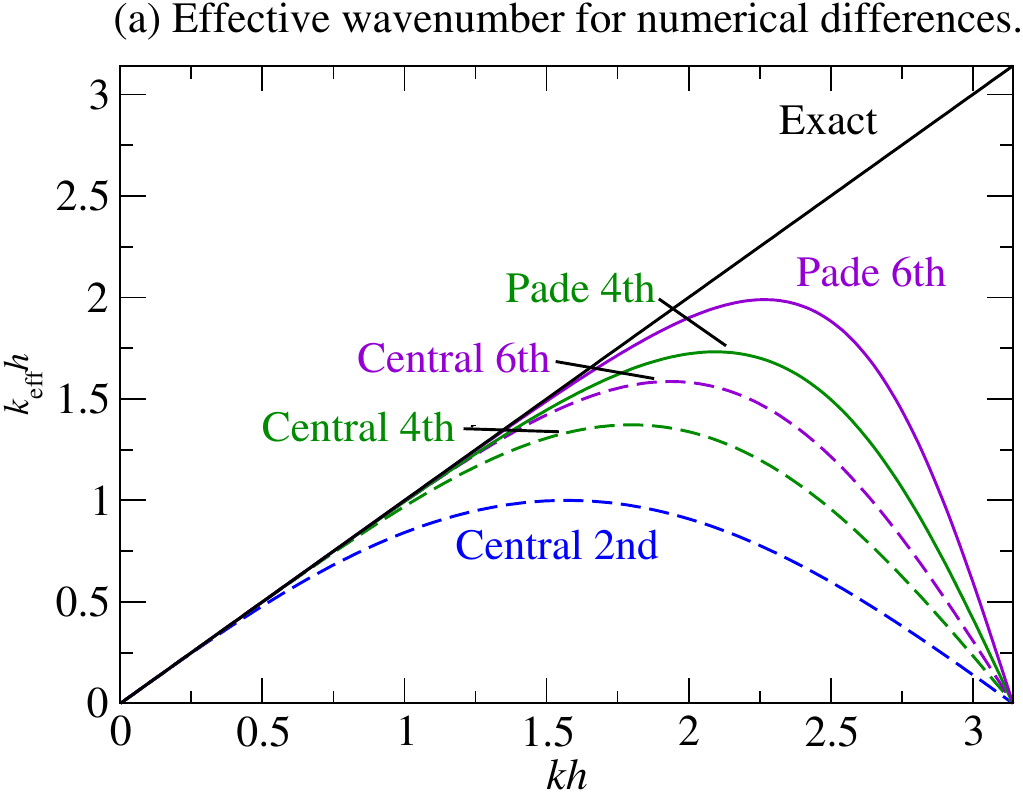} \hfill
\includegraphics[width=3.5in]{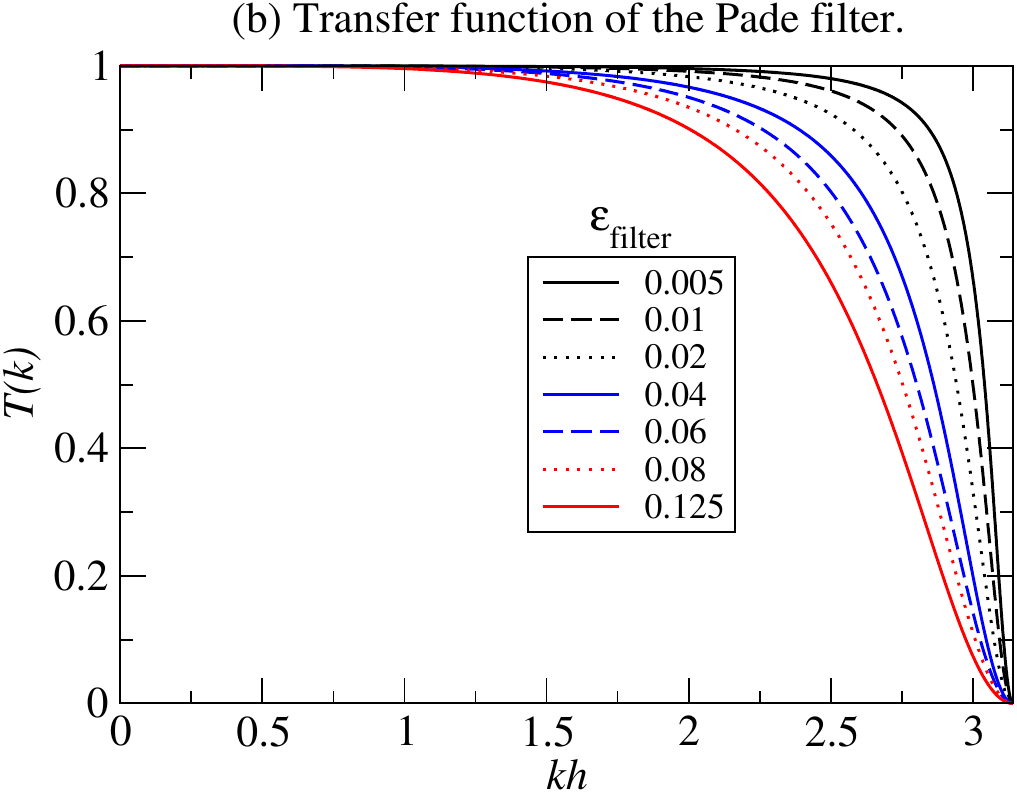}
\end{center}
\caption{(a) Effective wavenumber for various difference schemes. (b) Transfer function, $T(k)$, for the Pad\'e filter.  The grid spacing is $h$.  The limit $kh = \pi$ corresponds to the Nyquist mode which has wavelength $2h$.}
\label{fig:keff}
\end{figure}
Figure~\ref{fig:keff}a displays the effective wavenumber for various schemes and shows that the fourth-order Pad\'e scheme has better resolving power than a conventional sixth order scheme.  In general, $k_\rmeff(k)$ is complex and its imaginary part represents numerical dissipation of the scheme.  The fact that $k_\rmeff(k)$ is real for central schemes means that they have dispersion error but no dissipation error (for periodic boundary conditions).  As as example of how to use the effective wavenumber diagram, we estimate by eye that the highest wavenumber for which we can trust the 4th order Pad\'e scheme is $kh = 1.5$ which implies that the smallest number of grid points per wavelength for which the scheme is accurate is about four.

\subsection{Pad\'e filtering}\label{sec:filter}

It is known that in the presence of non-linearity or non-uniform grids, Pad\'e schemes produce small $2\Delta$ waves at every time step which can grow if not controlled; this is true for conventional central schemes as well.  Our remedy is to apply a minimum amount of Pad\'e filtering \citep[][\S C.2]{Lele_1992} which has a sharp cut-off and targets the very highest wave numbers.  We also use Pad\'e filtering as an implicit sub-grid treatment; this is discussed in \S\ref{sec:tests} where we use it for this purpose in an axisymmetric simulation of vertical shear instability (VSI).

 We use the fourth-order filter of \cite{Lele_1992} which for periodic boundary conditions, or in the interior of non-periodic domains, has the form
\be
a U_{j-1} + U_j + aU_{j+1} = P(u_{j-2}+u_{j+2}) + Q(u_{j-1}+u_{j+1}) + Ru_j, \eql{filter}
\ee
where $U_j$ are the filtered $u_j$.
The conditions for fourth-order accuracy are:
\be
a = -\frac{1}{2} + 2Q, \hskip 0.5truecm
R = \frac{1}{2}\left(2 + 3a - 3Q\right), \hskip 0.5truecm
P = \frac{1}{4}\left(a - Q\right).  \eql{fc} 
\ee
The reader can verify that there is no filtering when $Q = 1/2$.  To specify the strength of the filter, the code uses the parameter $\epsilon_\mathrm{filter}$ such that
\be
   Q = \frac{1}{2} - \frac{1}{4}\epsilon_\mathrm{filter}. \eql{q}
\ee
The transfer function $T(k)$ of the filter versus wavenumber can be obtained by substituting $u_j = e^{\rmi k x_j}$ and $U_j = T(k) e^{\rmi k x_j}$ (with $x_j = jh$) into \eqp{filter}.  Figure~\ref{fig:keff}b shows $T(k)$ for various values of $\epsfil$.  Axisymmetric simulations for $\epsfil$ values ranging from $0.01$ to $0.125$ will be presented in \S\ref{sec:vsi}.

For non-periodic boundaries, we use the boundary treatment developed by Alan Wray (private communication) that is conservative, i.e., it preserves
\be
   \sum_{j=1}^N u_j h_j.
\ee
The formulae applied at $j = 2$ and $N-1$ are
\begin{align}
   B U_1  + C U_2    + D U_3    &= E u_1  + F u_2    + G u_3    + H u_4, \\
   B U_N + C U_{N-1} + D U_{N-2} &= E u_N + F u_{N-1} + G u_{N-2} + H u_{N-3}.
\end{align}
with
\be
B = 2a,  \hskip 0.3truecm C = 1 + a,   \hskip 0.3truecm  D = a,   \hskip 0.3truecm E = \frac{1}{4}(7a + Q), 
\hskip 0.3truecm F = \frac{1}{4}(4 + 7a - 3Q), \hskip 0.3truecm G = \frac{1}{4}(a + 3Q), \hskip 0.3truecm
  H = \frac{1}{4}(a - Q)
\ee
The boundary values $u_1$ and $u_N$ are unchanged.  

The first item of Table IX in \cite{Lele_1992} lists the the leading order truncation error for the filter \eqp{filter}:
\be
   U(x) = \left[1 + \frac{1}{16} \left(1 - 2a\right) h^4 \partial^4/\partial x^4 + \cdots\right] u(x). \eql{filter_trunc}
\ee
Next consider the expression
\be
   u(x, t+\Delta t) = (1 + \nu_4 \Delta t \partial^4/\partial x^4) u(x), \eql{euler_step}
\ee
which represents application of an Euler step to
\be
   \frac{\partial u}{\partial t} = \nu_4 \frac{\partial^4 u}{\partial x^4}
\ee
Comparing \eqp{filter_trunc} and \eqp{euler_step} gives
\be
   \nu_4 = \frac{1}{16} (1 - 2a) \frac{h^4}{\Delta t}. \eql{nu}
\ee
Hence, to leading order, Pad\'e filtering corresponds to fourth-order hyperviscosity.  Substituting the first member of \eqp{fc} and \eqp{q} into \eqp{nu} gives
\be
   \nu_4 = \frac{1}{16} \frac{\epsilon_\mathrm{filter}}{\Delta t} h^4. \eql{nu2}
\ee
Equation \eqp{nu} implies that as the time step is reduced $\epsilon_\mathrm{filter}$ should be reduced.
It should be noted that the matrix associated with the Pad\'e filter becomes ill-conditioned for very small values of $\epsfil$ and a sufficiently large number of grid points.  To alleviate this, the filter is applied every $N_\mathrm{filter}$ time steps (rather than after every step) and with $\epsfil$ increased by $N_\mathrm{filter}$.  In that case one should replace $\Delta t$ with $\Delta t N_\mathrm{filter}$ in \eqp{nu2}.

\subsection{Boundary schemes and global conservation}\label{sec:nbc}

At the end points of a non-periodic direction ($j = 1$ and $N$), the scheme \eqp{std_pade} involves values and derivatives outside the domain.  Therefore, at $j = 1$ and $N$ we use the third-order one-sided scheme from equation (4.1.1) in \cite{Lele_1992}:
\begin{align}
\fpr_1 + \alpha_1 \fpr_{2} &= h^{-1} \left(a_1 f_1 + b_1 f_{2} + c_1 f_{3}\right),\\
\fpr_N + \alpha_1 \fpr_{N-1} &= -h^{-1} \left(a_1 f_N + b_1 f_{N-1} + c_1 f_{N-2}\right),
\end{align}
with
\be
   \alpha_1 = 2, \myspace a_1 = -15/16, \myspace b_1 = 2, \myspace\mathrm{and} \myspace c_1 = 1/2.
\ee
According to a theorem for hyperbolic initial boundary value problems, one can reduce the order of accuracy at the boundary by one without affecting the global order of accuracy \citep{Gustafsson_1981}.  The scheme used at the interior points $j \in [2,N-1]$ is \eqp{std_pade}.

We wish the finite-difference scheme to possess a discrete conservation property.  Specifically, we require that a discrete version of the Leibniz integral rule be satisfied: a numerical integral of the numerical derivative should reduce to a difference of boundary values.
The is achieved for a Pad\'e scheme as described in \citet[][\S4.2]{Lele_1992} and
\cite{Brady_and_Livescu_2019}.  Its application to the present scheme is described in Appendix~\ref{sec:cons}.  Briefly, the condition to be satisfied is that the row entries in columns 2 to $N-1$ of matrix $B$ must have a weighted sum of zero.  Here $B$ is the matrix representation of the right-hand-side of \eqp{std_pade}:
\be
   A \bff^\prime = \frac{1}{h} B \bff.
\ee
Appendix~\ref{sec:cons} shows that this condition is satisfied for the present boundary scheme.  An alternate way to obtain conservation is the ``reconstruction by primitive function'' trick referred to above.


\subsection{Data partitioning for parallelization}

Since an entire line of data along $x$ is needed to compute a Pad\'e derivative along $x$, for any direction $x$, we employ the \textit{pencil} data structure.  Each processor is assigned a pencil of data that can be thought of as a long brick with the long side along the direction of differentiation.  Most of the work is done with $z$-pencils (i.e., with $z$ as the long direction).  This work includes initialization, output, and, time integration sub-steps.
To compute $r$ derivatives, we perform a so-called ``transpose'' such that each processor also has $r$ pencils, and similarly for $\phi$ derivatives.  The flowfield arrays with $z$, $r$, and $\phi$ pencilling, respectively, are dimensioned as follows:

\begin{verbatim}
q          (sr:er, sphi:ephi, nz, ndof)         ! pencil along z
q_r_space  (sphi:ephi, sz_r:ez_r, ndof, nr)     ! pencil along r
q_phi_space(sr:er, sz_phi:ez_phi, ndof, nphi)   ! pencil along phi
\end{verbatim}

Here {\tt ndof} refers to the number of degrees of freedom (number of flow variables) at each grid-point and the other dimensions can be read, for example, as follows: {\tt sz\_r:ez\_r}---starting $z$ index to ending $z$ index for an $r$ pencil.  Partitioning and transpose routines were taken from Alan Wray's {\sc Stellarbox} code.

\section{Tests} \label{sec:tests}

\subsection{Convergence for two-dimensional advection}

Here we solve the equation for a scalar, $f(z, \phi, t)$, uniformly advecting in the $z$ and $\phi$ directions:
\be
   \frac{\p f}{\p t} + \frac{\p f}{\p z} + \frac{\p f}{\p \phi}= 0, \hskip 0.3truecm z \in [0, 2\pi), \hskip 0.1truecm \phi \in [0, 2\pi),
\ee
with periodic boundary conditions and the initial condition
\be
   f(z, \phi, 0) = 1 + \frac{1}{2}\sin 2z \sin 2\phi.
\ee
The CFL is fixed at unity and the number of grid points $n_z$ and $n_\phi$ is varied. 
\begin{figure}
\centering
\includegraphics[width=3.5truein]{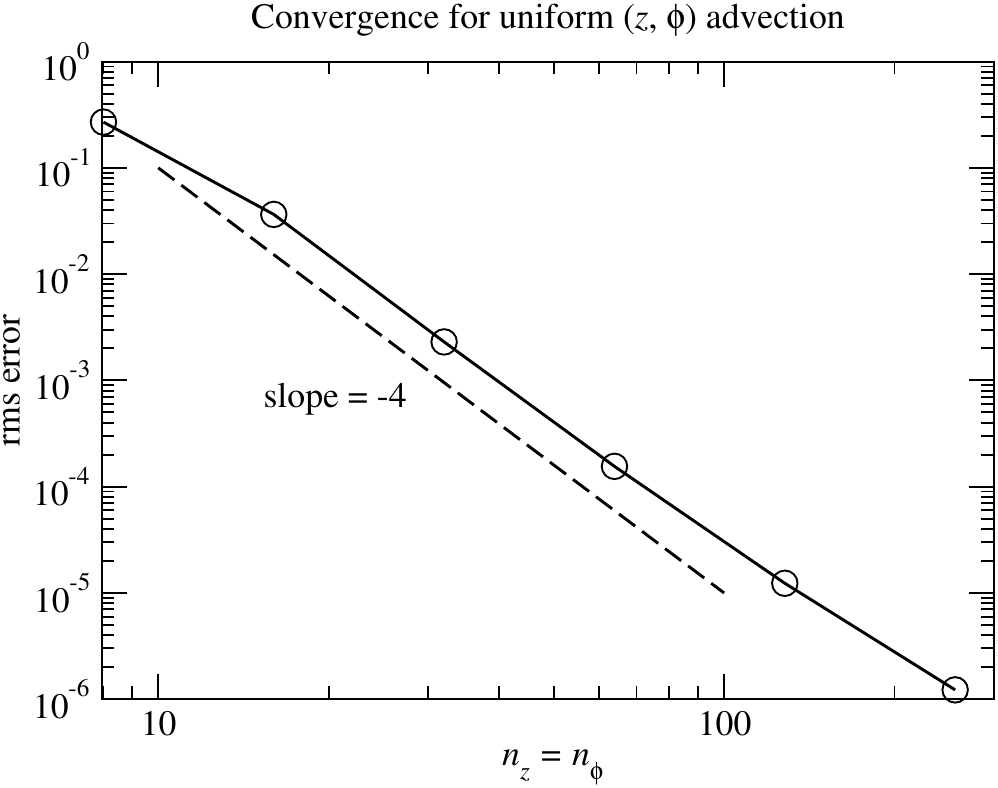}
\caption{Convergence for two-dimensional advection. \label{fig:advection_err}}
\end{figure}
Figure~\ref{fig:advection_err} plots the rms error at $t = 6\pi$ compared with the exact solution at $t = 6\pi$.  The rate of convergence is seen to be fourth-order.

\subsection{One-dimensional Euler equations with shocks}
\begin{figure}
\begin{center}
\vskip 0.25truecm
\includegraphics[width=3.5truein]{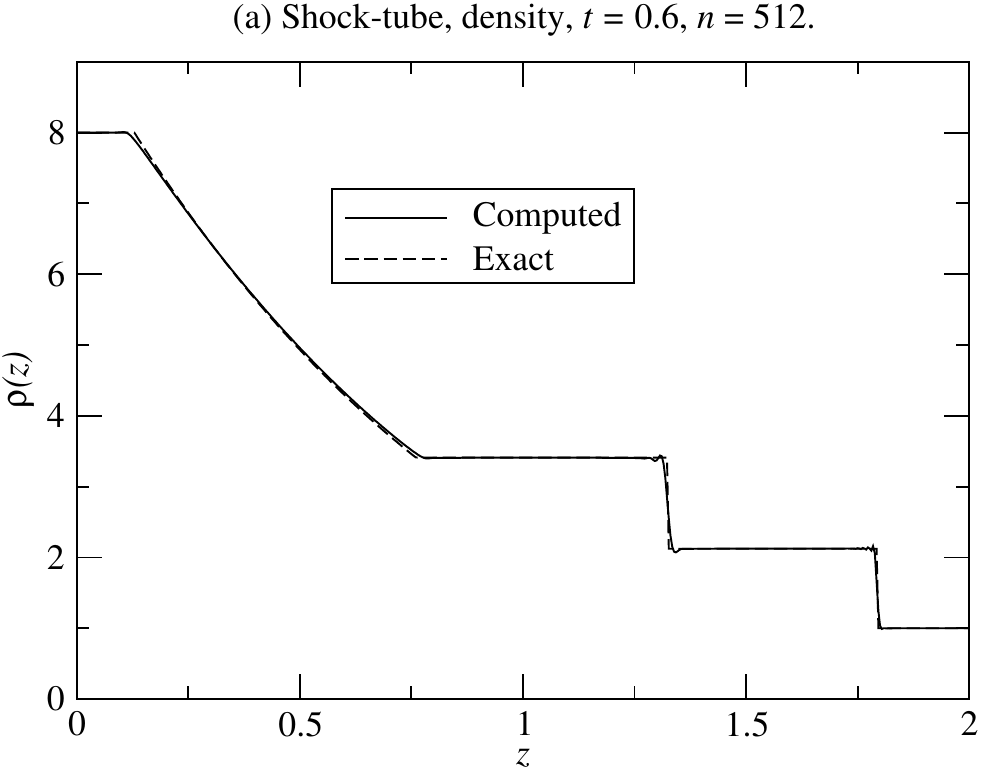}
\hfill
\includegraphics[width=3.5truein]{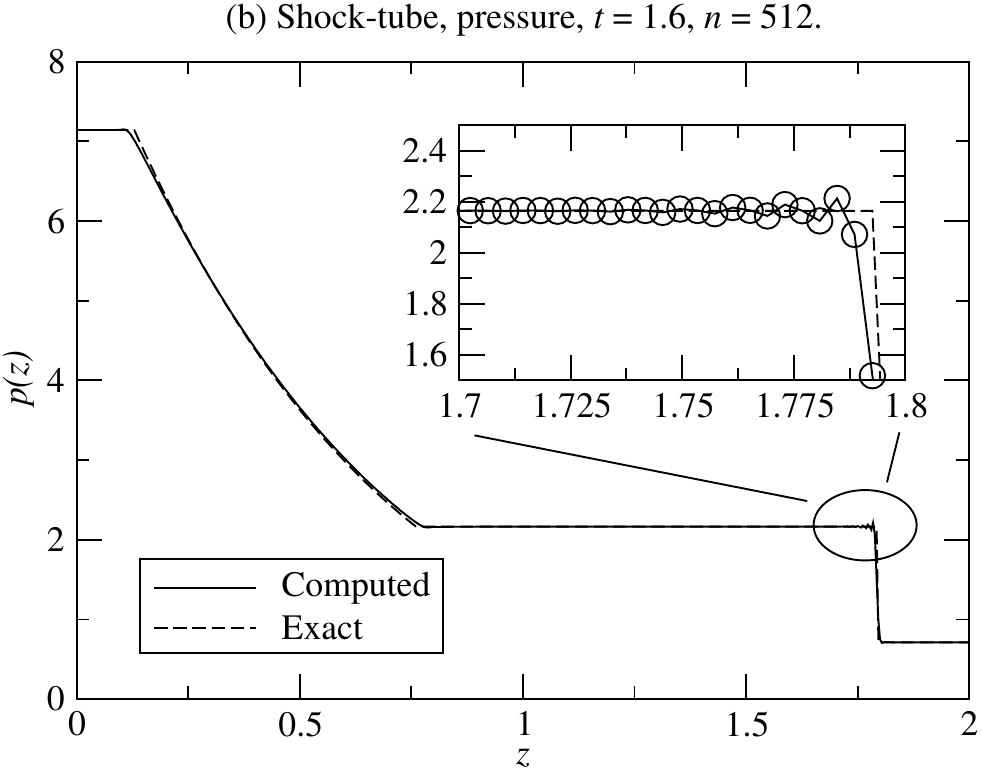}
\includegraphics[width=3.5truein]{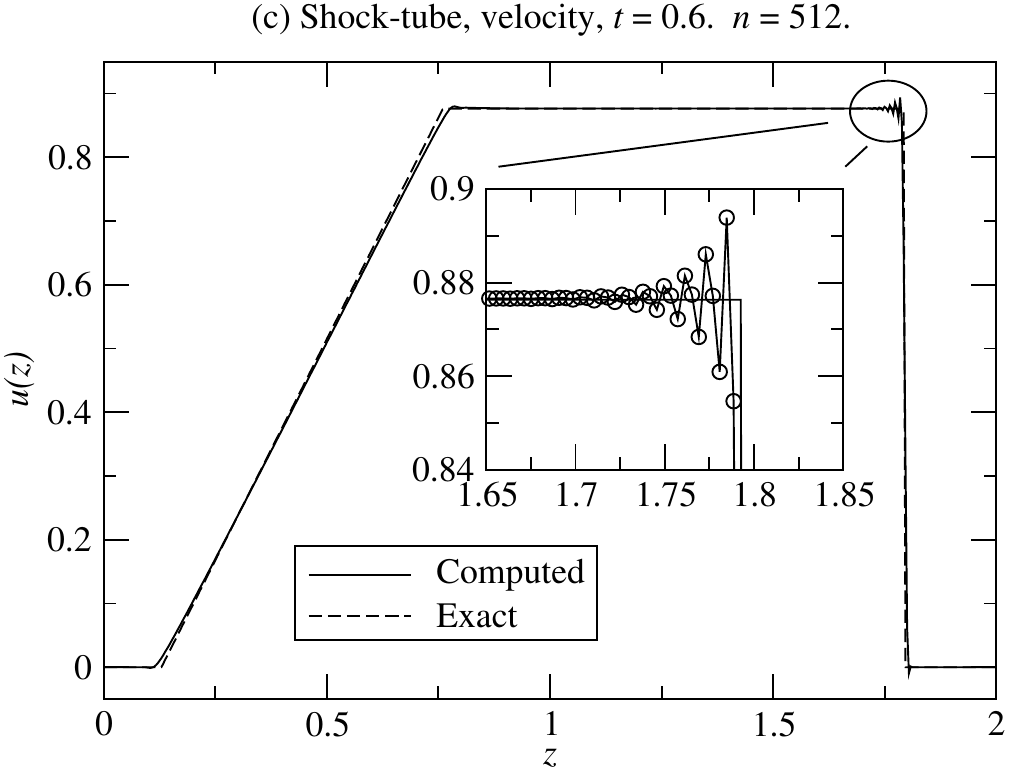}
\hfill
\includegraphics[width=3.5truein]{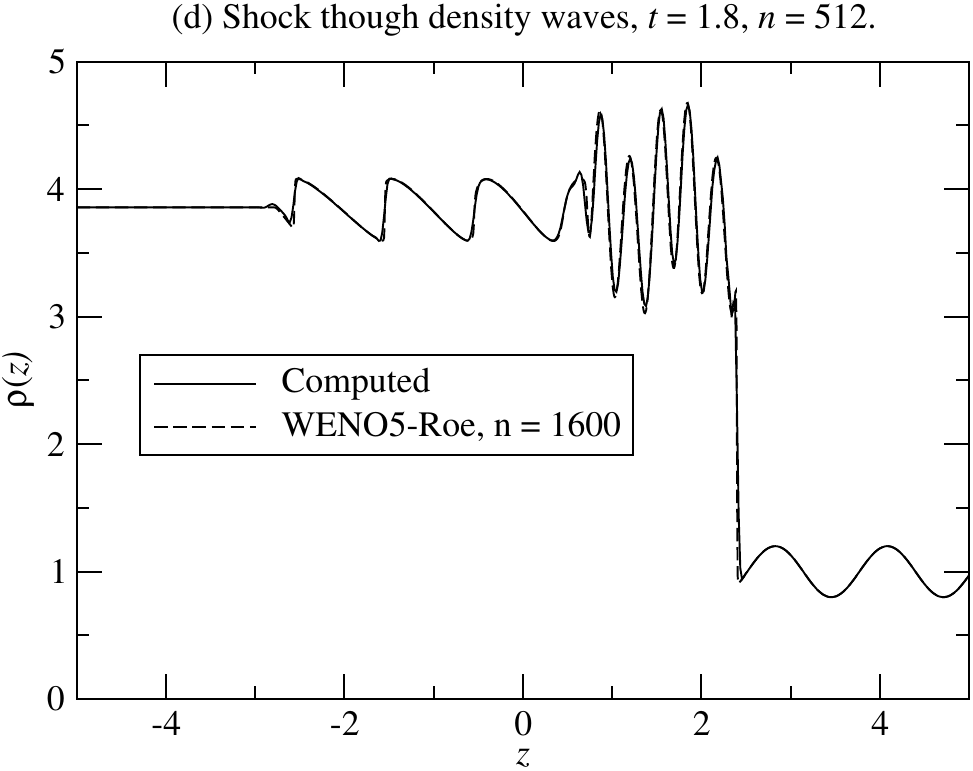}
\end{center}
\caption{1D Euler equation tests.  (a)--(c) : Shock tube. (a) Density.  (b) Pressure. (c) Velocity.  (d) Mach 3 shock propagating into density waves  \citep{Shu_and_Osher_1989}.  The result of the present code is compared with the WENO5-Roe method.  For both tests the coefficient of artificial pressure $\myCap = 2$ and 
Pade/compact filter strength $\epsfil = 0.05$ (applied at the end of every step). $\CFL = 1$.}
\label{fig:euler}
\end{figure}
Figure~\ref{fig:euler} presents two test cases for the one-dimensional Euler equations which was run in the code's $z$ direction by suppressing the other two.  Both tests employed $n_z = 512$ grid points and artificial pressure (with $\myCap = 2$) to capture shocks.  Figure~\ref{fig:euler}a shows the solution (solid line) to the 1D shock-tube problem with initial conditions to the left and right of the diaphragm (located at $x = 0.8$) as follows:
\be
(\rhoL, \uL, \pL) = (8, 0, 10/\gamma), \hskip 0.5truecm
(\rhoR, \uR, \pR) = (1, 0, 1/\gamma),
\ee
with zero velocity everywhere and $\gamma = 1.4$.
The computed solution is accurate; however, small oscillations are present in the post-shock region

\cite{Shu_and_Osher_1989} introduced the problem of a Mach 3 shock propagating through density waves as way of testing a method's ability to both capture shocks and resolve non-shock wavy features without excessive dissipation.  The initial condition is:
\be
(\rho, u, p) = \begin{cases}
(3.857143, 2.629369, 10.33333), & x < 4;\\
(1 + 0.2\sin 5z, 0, 1), & x \ge 4. \end{cases}
\ee
The result of this test is shown in Figure~\ref{fig:euler} where the baseline comparison (dashed line) was obtained using a fifth-order WENO scheme with $n_z = 1600$ points, the Roe flux, and reconstruction along characteristics.  The present code performs very well.

\subsection{Kelvin-Helmholtz instability}\label{sec:kh}

Here we consider the benchmark for viscous Kelvin-Helmholtz instability with a density gradient starting with smooth initial conditions constructed by \citet[][hereafter L2016]{Lecoanet_etal_2016}.  The full domain size is $L_x \times L_z = 1 \times 2$, however, only the lower half  of the $z$ domain will be shown, since the rest is shift-symmetric.  The resolution is $n_x \times n_z$ with $n_z = 2 n_x$.  It should be noted that L2016 write the heat conductivity as $k = \rho\chi$, where $\chi$ is the heat diffusivity.  The actual definition (which {\sc Pad\'e} uses) is $k = \rho \chi c_p$.  Therefore, to match L2016 we needed to divide our $k$ by $c_p$ which equals $5/2$ for the set-up of $L1016$, which assumes that the gas constant $R = c_p - c_v = 1$ and $\gamma \equiv c_p/c_v = 5/3$.  We chose time step $\Delta t$ so that the Courant-Friedrichs-Lewy number was 1.5.  We initially applied the Pad\'e filter after every step with $\epsfil = 0.0025$, however, the associated matrix becomes ill-conditioned for very small values of $\epsfil$ when the number of grid points is sufficiently large.  To alleviate this, the strength of the Pad\'e filter was applied every 40 time steps with $\epsfil = 0.10$.  The simulation was run on a laptop with an Apple M2 Pro chip using 8 cpus.  The cpu time per step was 0.3 s.

\begin{figure}
\centering
\includegraphics[width=3.4in, trim=3 3 3 3,clip]{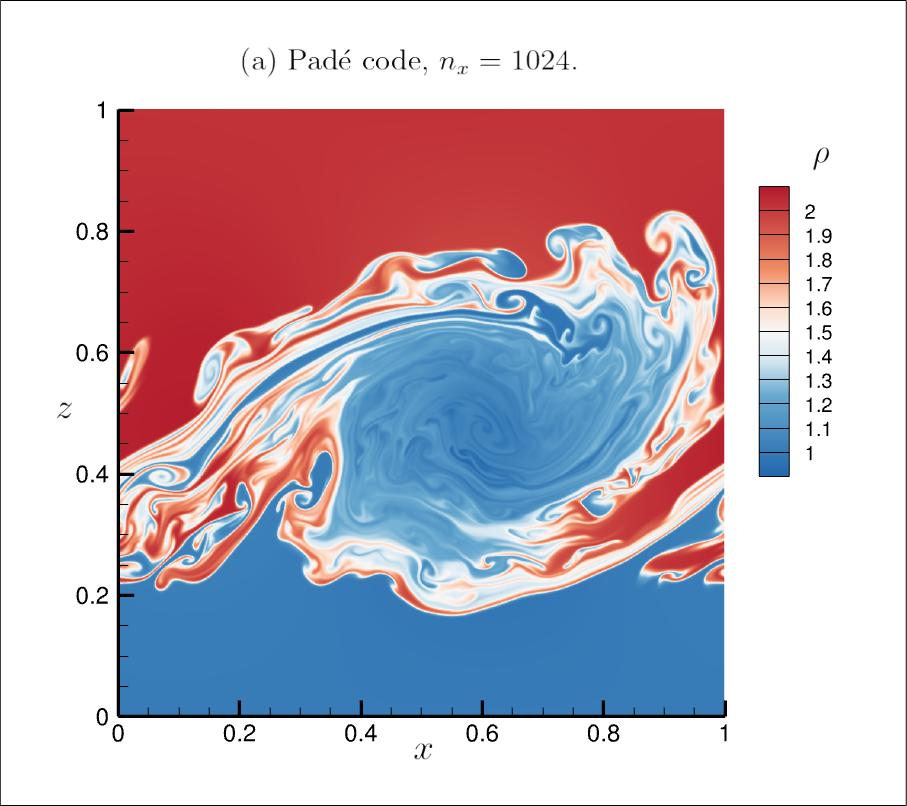} \hfill
\includegraphics[width=3.4in, trim=3 3 3 3,clip]{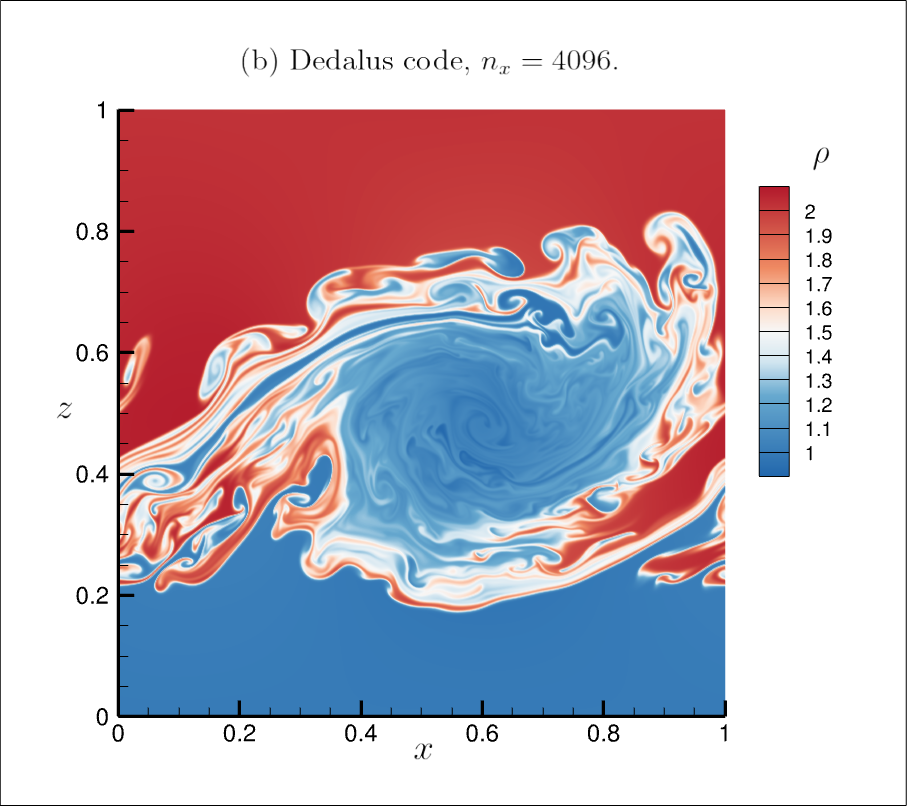} \hfill
\caption{Density field at $t = 6$ for the Kelvin-Helmholtz instability benchmark of \cite{Lecoanet_etal_2016}.  (a) {\sc Pad\'e} code with $n_x = 1024$. (b) Pseudo-spectral {\sc Dedalus} code with $n_x = 4096$.  Data provided by D. Lecoanet.}
\label{fig:kh}
\end{figure}
Figure~\ref{fig:kh}a shows the density field at $t = 6$ obtained from {\sc Pad\'e} with $n_x = 1024$.  It is compared with the result of L2016 obtained using their pseudo spectral code {\sc Dedalus} with $n_x = 4096$; we thank Prof. D. Lecoanet (Northwestern Univ.) for sending us the data.  The agreement is very good.

\subsection{Taylor-Couette flow}

A natural test case in cylindrical coordinates is Taylor-Couette flow, i.e., viscous flow driven by rotating inner and outer cylinders with radii $\rinner$ and $\router$, respectively.  The corresponding rotation rates are $\Omegai$ and $\Omegao$ giving corresponding rotational speeds $U_\rmi \equiv \Omega_\rmi r_\rmi$ and $U_\rmo \equiv \Omega_\rmo r_\rmo$.  Two of the four non-dimensional parameters are the inner and outer Reynolds numbers $\Rei \equiv U_\rmi d/\nu$ and
$\Reo \equiv U_\rmo \rinner d/\nu$, where $d = \router - \rinner$ is the gap width.  The third and fourth non-dimensional parameters are the ratio $\eta \equiv \rinner/\router$, and the non-dimensional vertical domain size $\lambda \equiv L_z/d$.

The present code solves the compressible equations while most simulations reported in the literature are for incompressible flow.  To approximate incompressible simulations the rotational Mach number of the inner cylinder is set to $0.1$, the equation of state is isothermal, and isothermal boundary conditions are applied at the two walls.  The initial density is set to a uniform value $\rho_0$.  Code units are such that $U_\rmi = d = \rho_0 = 1$.

Torques, $G_\rmi$ and $G_\rmo$, per unit axial length exerted on the fluid by the inner and outer cylinders, respectively, are computed as diagnostics.
\be
   \mathrm{torque} = \mathrm{area\ averaged\ shear\ stress} \times r \times \mathrm{area}.
\ee
The area averaged shear stress is given by
\be
   \tau_{r\phi} = \left<\mu r \frac{\p}{\p r}\left(\frac{u_\phi}{r}\right)\right>,
\ee
where $\mu$ is the dynamic viscosity and $\left<.\right>$ denotes an average over the surface.
Therefore,
\begin{align}
   G_\rmi &= -2 \pi r_{\rmi}{^3} \left<\frac{\p}{\p r}\left(\mu\frac{u_\theta}{r}\right)\right>_{r = r_\rmi}, \eql{Gi}\\
   G_\rmo &=+2 \pi r_{\rmo}{^3} \left<\frac{\p}{\p r}\left(\mu\frac{u_\theta}{r}\right)\right>_{r = r_\rmo}. \eql{Go}
\end{align}
The negative sign in \eqp{Gi} arises from the fact that at $r = r_\rmi$, the normal to the fluid surface is in the $-r$ direction.
For steady flow, conservation of angular momentum implies that $G_\rmi$ and $G_\rmo$ must be equal and opposite in sign.
\begin{table*}
\begin{center}
\begin{tabular}{c c c c c c c}
\toprule
 $\eta \equiv r_\rmi/r_\rmo$  & $L_z/d$   & $\Rei$   & $G_\rmi$    &  $G_\rmo$ & Published $G_\rmi$ & Ref. \\
 \midrule
 0.875                                    & 2.5        & 139.22          & 3.3485       & -3.3482     & 3.3539 & \cite{Marcus_1984}\\
 0.50                                      & 1.988    & 78.8               & 1485        & -1485       & 1487   & \cite{Moser_etal_1983}\\
\bottomrule
\end{tabular}\end{center}
\caption{Computed torques  $G_\rmi$ and $G_\rmo$ per unit axial length exerted on the fluid by the inner and outer cylinders, respectively, for steady axisymmetric Taylor-Couette flow.  These values are compared with values obtained in the cited references.  The values are normalized differently in the two references as described in the text.}
\label{tab:torque}\end{table*}
We first consider two axisymmetric cases that produce a steady flow with counter-rotating vortices.  The inner cylinder rotates at angular velocity $\Omega_\rmi$ while the outer cylinder is fixed.  Each case was run using a $32 \times 32$ grid ($n_r \times n_z$) with $\epsfil = 0.005$.  The first two entries of Table~\ref{tab:torque} shows that the computed torque per unit length agrees with previous published results.  \cite{Marcus_1984} uses units in which $\rho_0 = d = \Omega r_\rmi = 1$.  \cite{Moser_etal_1983} use the same units but normalize $G$ by $\rho_0 \nu^2$ where $\nu = \mu/\rho_0$.  The values in Table~\ref{tab:torque} use the same conventions.

Finally, a case of 3D unsteady counter-rotating Taylor-Couette flow is considered following \cite{Dong_2008}.  The inner and outer cylinders rotate counter-clockwise and clockwise, respectively with $\Rei = -\Reo = 500$, $\eta = 0.5$ and $L_z/d = 2\pi$.  \cite{Dong_2008} defines the non-dimensional torque coefficient for the inner cylinder as
\be
\left(C_\mathrm{T}\right)_\rmi = \frac{G_\rmi}{\frac{1}{2} \pi \rho_0 U_\rmi^2 r_\rmi^2 L_z},
\ee
and similarly for the outer cylinder.  The number of grid points is $48 \times 96^2$ ($n_r, n_z, n_\phi$) and the strength of the Pad\'e filter was set to $\epsilon_\mathrm{filter} = 0.005$.
\begin{figure}
\begin{center}
\includegraphics[width=1.5in, trim=2 2 2 2,clip]{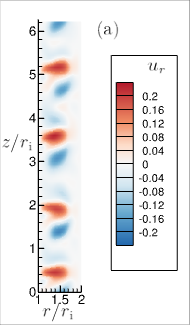}\hfill
\includegraphics[width=2.4in, trim=2 2 2 2,clip]{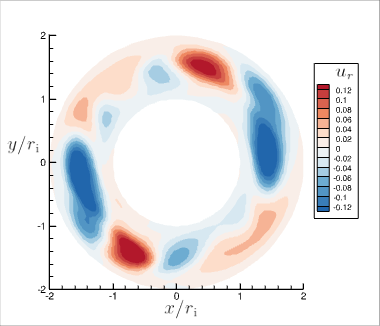}\hfill
\includegraphics[width=2.9in]{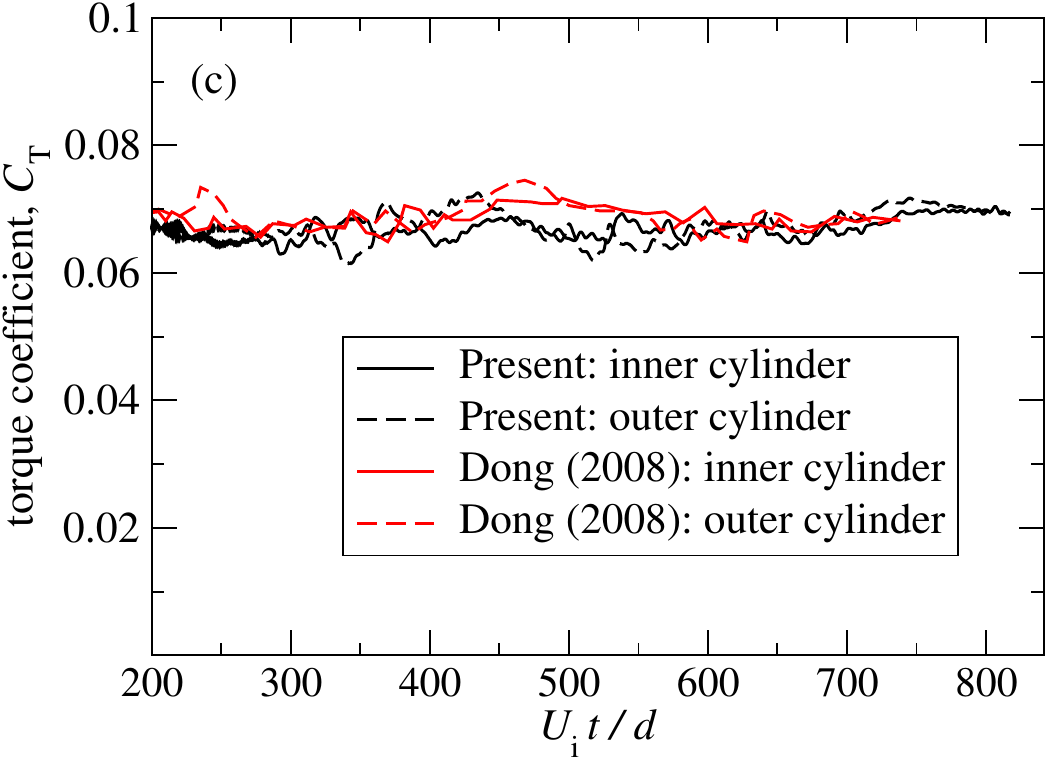}
\end{center}
\caption{Counter-rotating Taylor-Couette flow at $\Rei = -\Reo = 500$ following \cite{Dong_2008}. The mesh is $48 \times 96^2 (n_r \times n_z \times n_\phi)$.  (a) \& (b) Radial velocity contours at $U_\rmi t/d = 798.8$.   (a) The meridional plane $\phi = 0$. (b) Horizontal plane at mid-height ($z = \pi$) to show that the flow is three-dimensional.  (c) Torque coefficients (black) compared with curves (red) digitized from \citet[][Figure 3]{Dong_2008} and shifted backward in time by $U_\rmi\Delta t/d = 802$.  The negative of the torque coefficient is plotted for the outer cylinder.}
\label{fig:tc}
\end{figure}
Figures~\ref{fig:tc}a and b shows the radial velocity in a meridional and horizontal plane and reveals the three-dimensionality of the flow.  Figure~\ref{fig:tc}c shows the torque coefficients for the inner and outer cylinders after the flow has reached statistical stationarity.  The values agree with those shown in Figure 3 of \cite{Dong_2008}.

\subsection{Vertical shear instability and the effect of varying the strength of the Pad\'e filter}\label{sec:vsi}

Here we present results for axisymmetric vertical shear instability (VSI) at an early stage of evolution following the set up of \cite{Nelson_etal_2013}.  Detailed results will be presented in a forthcoming publication which will also include 3D results.  
%
\begin{table}
\centering
\begin{tabular}{l c c c c}
\toprule
 Parameter & Value\\
\midrule
Orbital period, $T_0$, at $r_0$ & $1$\\
Density, $\rho_0$, at $r_0$ & $1$\\
Scale height, $H_0$, at $r_0$ &$1$\\
Density exponent, $p$ &$-3/2$\\
Temperature exponent, $q$               &$-1$\\
Disk aspect ratio, $H_0/r_0$ & $0.10$\\
Radial domain (including sponge), $[\rmin/H_0, \rmax/H_0]$ & $[6.5, 13.5]$\\
Vertical domain (including sponge), $[\zmin/H_0, \zmax/H_0]$ & $[-3.5, 3.5]$\\
Width of sponge at domain border, $\delta_\mathrm{sponge}/H_0$ & 0.5\\
Decay period, $t_\mathrm{sponge}$, for sponge, & 20 time steps\\
Number of grid points, $n_r \times n_z$ & $512^2$\\
Strength of Pad\'e filter, $\epsilon_\mathrm{filter}$ & 0.01-0.125\\
\bottomrule
\end{tabular}
\caption{Parameters for the axisymmetric vertical shear instability run.  $H_0$ is the disk scale height at mid-radius and was set to unity.  Subscript `0' refers to a quantity evaluated at mid-radius.}
\label{tab:vsi}
\end{table}
The parameters of the set-up are given in Table~\ref{tab:vsi}.  A locally isothermal equation of state
\be
   p = \rho c_\mathrm{i}^2(r),
\ee
is used which represents the case of infinitely rapid relaxation of temperature to the basic state.  
The square of the sound speed, which is proportional to the temperature, is specified as a power-law:
\be
   c_\mathrm{i}^2(r) = c_0^2 \left(r / r_0\right)^q,
\ee
where the temperature exponent $q = -1$, $r_0$ is the mid-radius of the computational domain, and $c_0$ is chosen to make the scale height $H_0 \equiv H(r_0) = 1$.

Zero normal velocity boundary conditions are applied at all four domain edges.  Each edge abuts a sponge region of width $\delta_\mathrm{sponge} = 0.5 H_0$ where the flow relaxes back to the basic state with a characteristic period $t_\mathrm{sponge}$ given in the table.
\begin{figure}
 %
\hskip 0.2truecm             
\centerline{
\includegraphics[width=3.5truein, trim=2 2 2 2,clip]{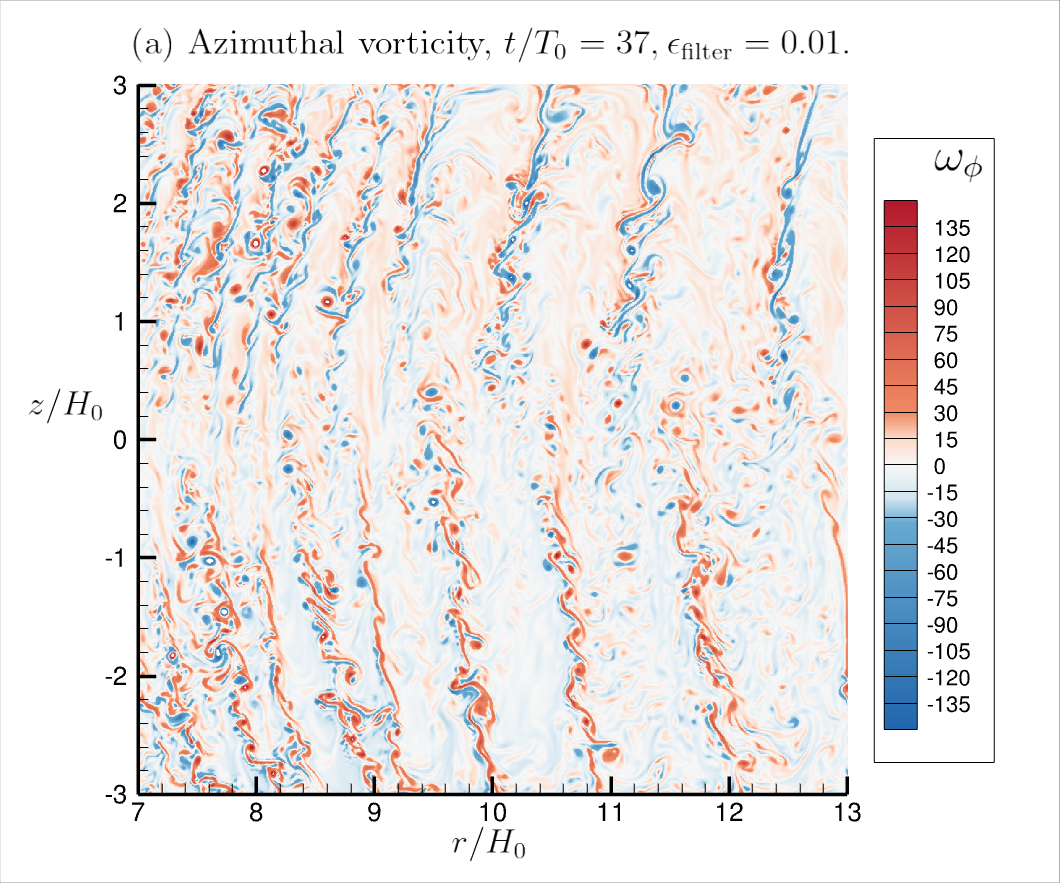} \hfill
\includegraphics[width=3.5truein, trim=2 2 2 2,clip]{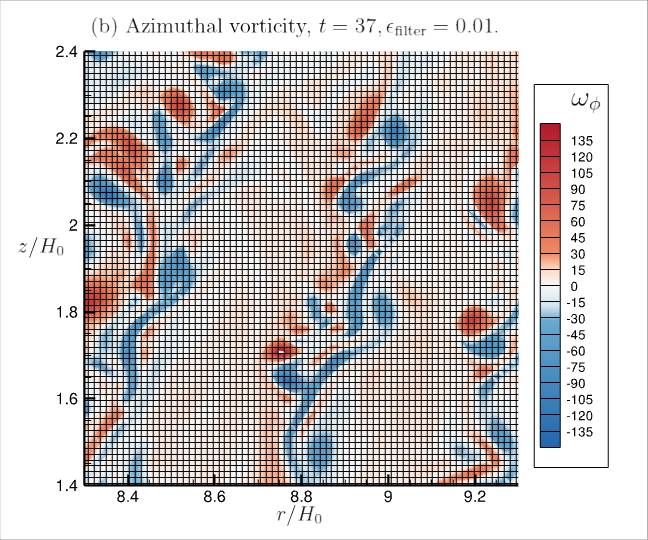}
                } \vskip 0.2truein
\centerline{
\includegraphics[width=3.5truein, trim=2 2 2 2,clip]{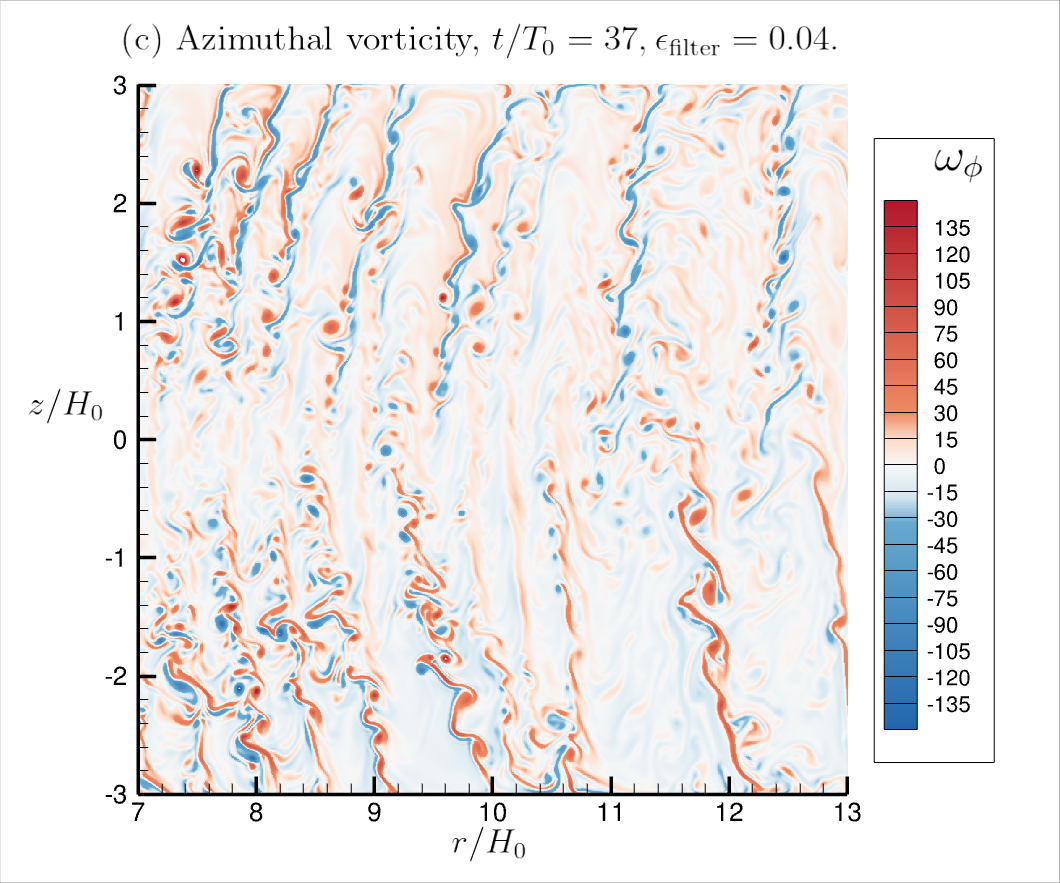} \hfill
\includegraphics[width=3.5truein, trim=2 2 2 2,clip]{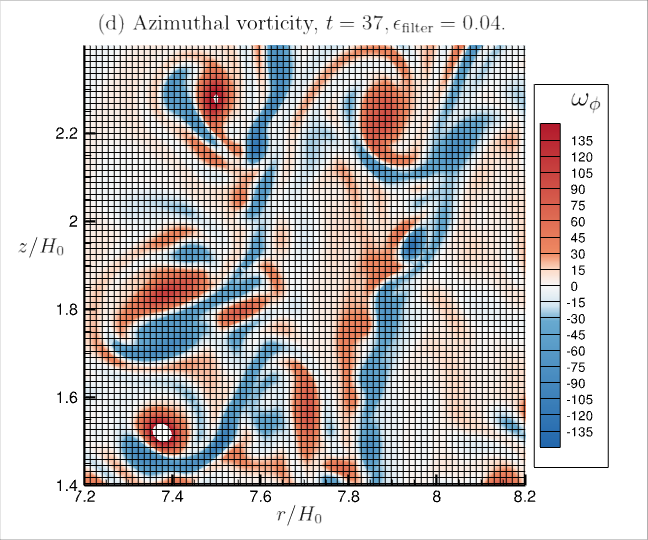}
                }  \vskip 0.2truecm              
\centerline{
\includegraphics[width=3.5truein, trim=2 2 2 2,clip]{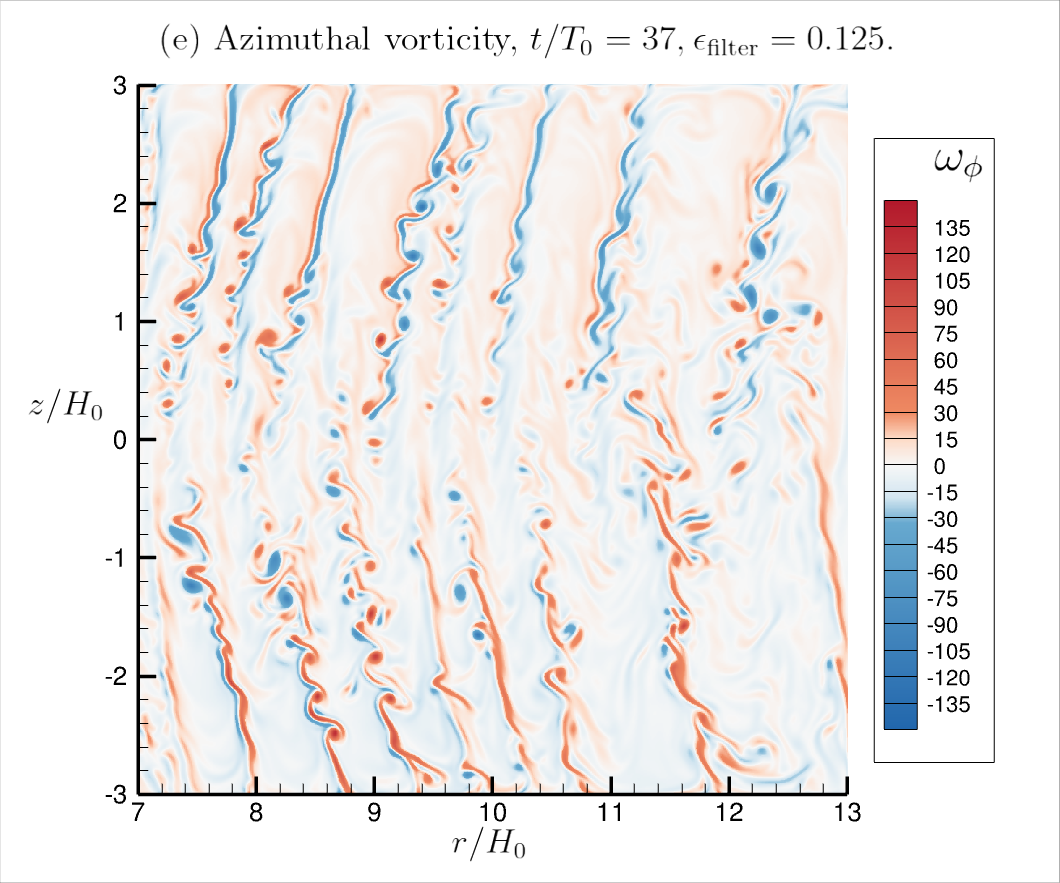} \hfill
\includegraphics[width=3.5truein, trim=2 2 2 2,clip]{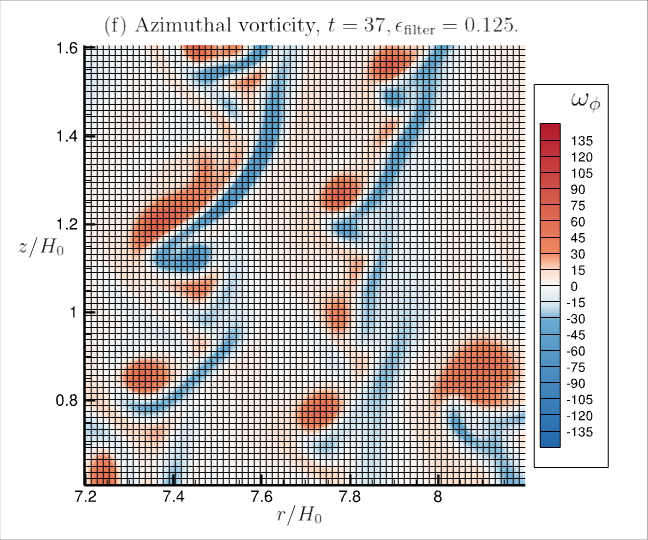}
                } \vskip 0.2truecm
\caption{Effect of varying the strength of the Pad\'e filter, $\epsfil$ when the Pad\'e filter is used as an implicit sub-grid scale treatment for an axisymmetric VSI simulation.  Completely white pixels are where $\omega_\phi$ exceeds the range of the legend.  The mesh size is $512^2$.  The plots in the right-hand column are intended to show the level of numerical $2\Delta$ oscillations which appear as sawtooth features.}
\label{fig:vary_eps}
\end{figure}

Figure~\ref{fig:vary_eps} shows the azimuthal vorticity, $\omega_\phi$ at $t/T_0 = 37$ which is just after saturation of the linear phase of the instability.  We caution the reader that the statistically stationary state is quite different and reached much later at $t/T_0 \approx 350$; this will be reported in a later publication.  The result of using three different values for the strength, $\epsfil$, of the Pad\'e filter is shown.  
\begin{figure}
%
\centering
\includegraphics[width=3.4in, trim=2 2 2 2,clip]{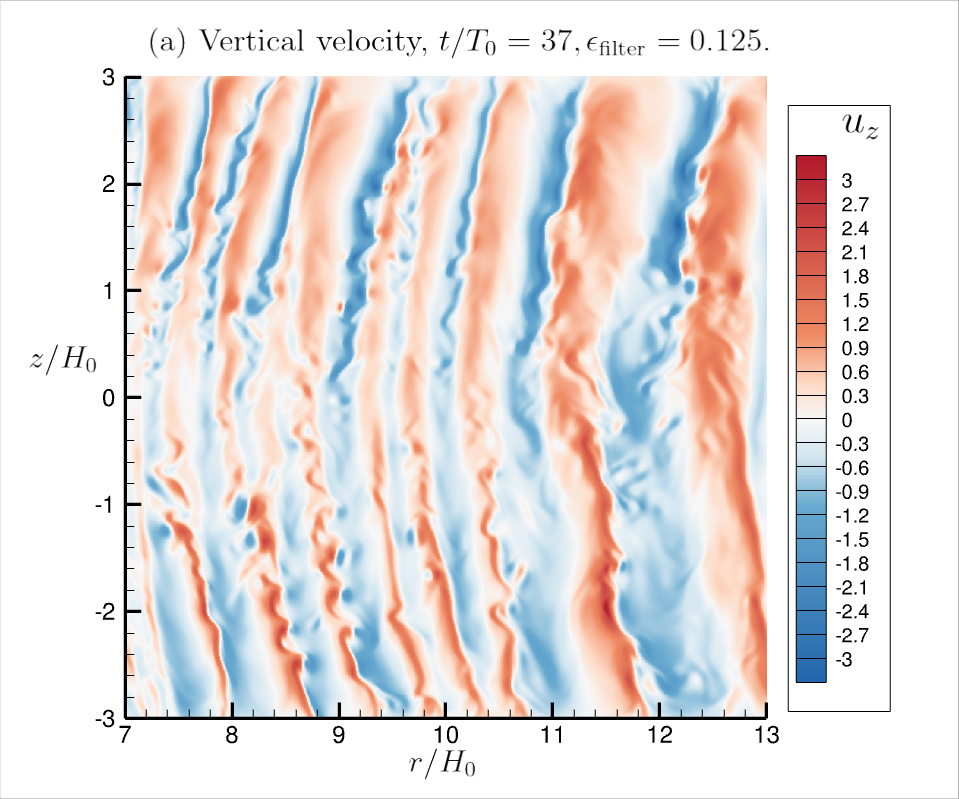} \hfill
\includegraphics[width=3.4in, trim=2 2 2 2,clip]{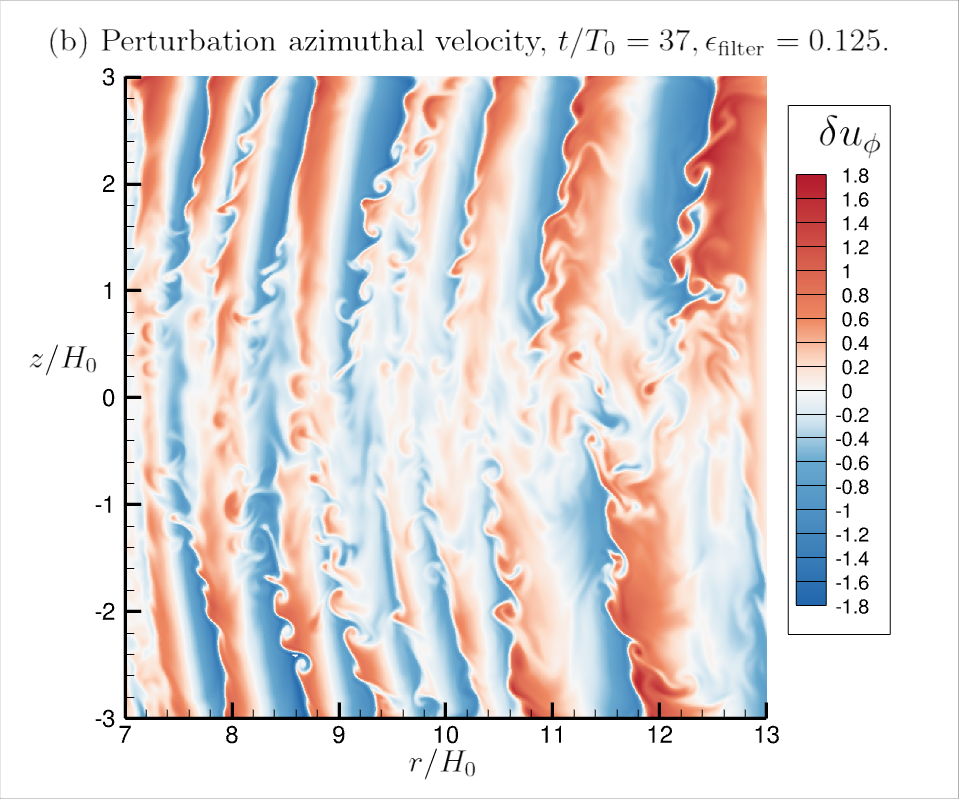} \hfill
\caption{(a) Vertical velocity for the axisymmetric vertical shear instability simulation using a $512^2$ mesh.  The units of velocity are scale heights ($H_0$) per orbit time ($T_0$). (b) Similarly, the perturbation azimuthal velocity.}
 \label{fig:uz}
\end{figure}
The azimuthal vorticity consists of pairs of shear layers of opposite sign which induce up and down jets of vertical velocity (Figure \ref{fig:uz}a) that are characteristic of VSI \citep{Nelson_etal_2013}.  The shear layers roll up into discrete eddies via the Kelvin-Helmholtz instability.  It is important to note that VSI also produces shear layers with $\p \delta u_\phi / \p r$ shear, i.e., radial gradients of perturbation angular velocity.  These are shown in Figure \ref{fig:uz}b which plots the azimuthal velocity perturbation, $\delta u_\phi$, and shows that is it about $2/3$ the value of the vertical velocity perturbation.  This shear will also be subject to Kelvin-Helmholtz instability, however, it will be modified by the presence of mean Keplerian shear.

We now discuss how a user should choose the filter strength, $\epsfil$.  Direct numerical simulations (DNS) of turbulent flow are performed with molecular viscosity and all the scales of the turbulence down to the dissipation scale are well resolved.  In this case, the very minimum value of $\epsfil$ should be chosen.  For instance in the Taylor-Couette simulations we chose $\epsilon_\mathrm{filter} = 0.005$.  

However, the Reynolds number in protoplanetary disks is too large for numerical simulations to be able to resolve all the turbulent scales.  Therefore, some treatment of the unresolved scales is required.  For engineering and geophysical flows, the most common practice is to use an explicit model for the sub-grid stresses, the simplest being the  Smagorinsky model.  Another approach, followed for all protoplanetary disk simulations to date, is to simply let the dissipation inherent in the numerical method damp scales near the grid cut-off.  This procedure is referred to as implicit large-eddy simulation (ILES) and was first articulated by \cite{Boris_etal_1992}.  Comparison with direct numerical simulations (DNS, in which all scales are resolved) have since shown that it is accurate \citep{Ritos_etal_2018}.  In our approach, we use the dissipation provided by the Pad\'e filter as an implicit sub-grid treatment.  To leading order, the Pad\'e filter corresponds to a fourth-order hyperviscosity; see equation (30) in \citet{Shariff_and_Wray_2018} and the discussion following it.  

When the Pad\'e filter is used an ILES treatment, $\epsfil$ should be chosen to balance the desire to capture as wide a range of small scales possible (with a fixed grid size) by lowering $\epsfil$, while at the same not allowing to much energy to to pile-up in $2\Delta$ waves.   For illustration, Figure~\ref{fig:vary_eps} shows the result of varying the strength, $\epsfil$, of the filter on the vorticity field.  The reader may refer back to Figure~\ref{fig:keff}b which shows the filter transfer function for different $\epsfil$ choices.  The left hand column of plots in Figure~\ref{fig:vary_eps} shows that with reduced $\epsfil$ more finer scales of the flow are resolved.  The right-hand column of plots zoom in to a square region with sides equal to $H$.  For the lowest filter strength ($\epsfil =0.01$), $2\Delta$ (sawtooth) oscillations can be observed in thin vortex layers oriented at $45^\circ$ to the mesh.  For $\epsfil = 0.04$ and 0.125, smaller amplitude oscillations are present in one vorticity layer whose width is about one grid diagonal, which is very thin indeed. These oscillations are not visually detectable in plots of the velocity field even for $\epsfil = 0.01$.  We conclude that $0.125$ would be a conservative choice for $\epsfil$.

Finally, we would like to discuss some physical effects that manifest as diffusivity represented by the filter is reduced and the effective resolution is increased.  (a) The rolled-up vortices are smaller and there are more of them.  This is explained as follows.  The shear-layer thickness, $\delta$, is reduced with smaller diffusivity.  The most amplified Kelvin-Helmholtz (KH) wavelength $\lambda \approx 2\pi\delta$ is therefore also reduced, and with it the size of the vortices.  The number of vortices increases because there are more waves per unit length.  (b) The shear-layer vorticity increases.  We have that $\omega_\phi \sim \Delta U / \delta$ where $\Delta U$, the jump in velocity across each shear layer, is independent of diffusivity.  Therefore a reduction in $\delta$ with diffusivity leads to increased $\omega_\phi$. (c) The rolled-up vortices appear earlier.  The KH growth-rate $\propto \Delta U k_\rmmax$, where $k_\rmmax$ is the most amplified wavenumber. Since $k_\rmmax$ increases with reduced thickness, the KH vortices develop earlier with reduced diffusivity. (d) The vorticity in the the rolled-up vortex cores increases.  For each vortex we have $\omega_\phi \sim \Gamma / \mathrm{area}$, where $\Gamma \sim \Delta U \lambda$ is its circulation.  Since the vortex area $\sim \lambda^2$, we get that $\omega_\phi \sim \Delta U / \lambda$ which increases since $\lambda$ decreases.

\subsection{3D vertical shear instability: comparison of Fargo and non-Fargo runs}\label{sec:vsi3d}

In \citet{Shariff_and_Wray_2018} a comparison (with plots of error) was made between runs with and without the Fargo treatment for the case of two co-rotating vortices in a razor thin disk.
Here, we perform a similar comparison for 3D vertical shear instability (VSI).  The simulation was first run till $t/T_0 = 300.27$ with Fargo activated.  Next runs were made with and without Fargo for one orbital period ($T_0$) at mid-radius.  The run  parameters are the same as for the axisymmetric run presented in Table~\ref{tab:vsi}.  The only differences are a resolution of $512 \times 512 \times 1024$ ($n_r \times n_z \times n_\phi$) with an azimuthal domain of $\phi \in [0, 2\pi)$, and a Pad\'e filter strength of $\epsfil = 0.125$.  For the non-Fargo run, the Pad\'e filter was applied every other time step due to the fact that this run required about twice as many time steps as the run with Fargo.
\begin{figure}
\centering
\includegraphics[width=3.4in, trim=2 2 2 2,clip]{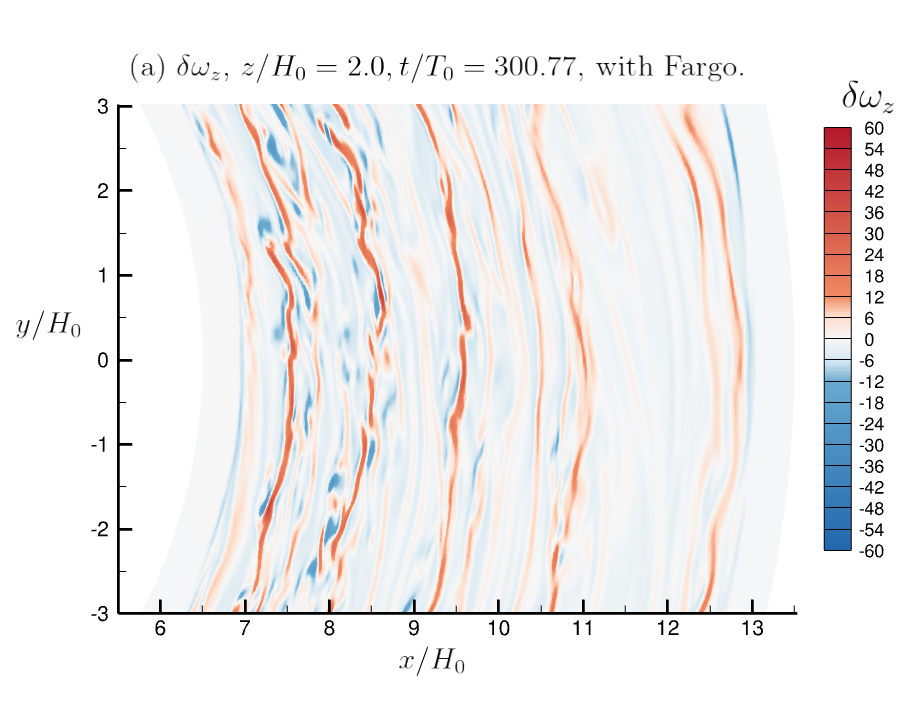} \hfill
\includegraphics[width=3.4in, trim=2 2 2 2,clip]{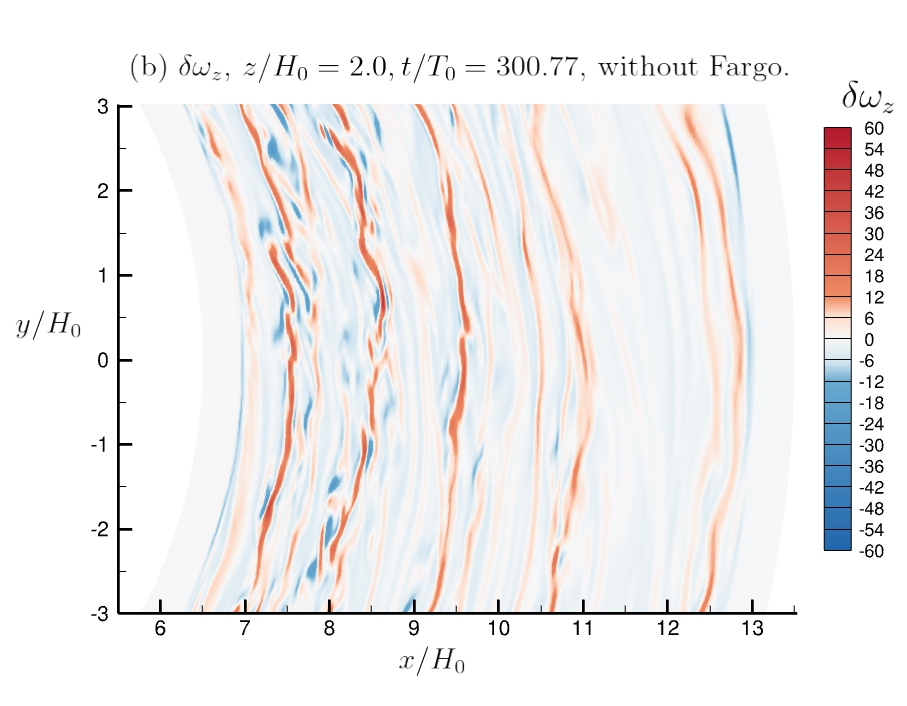} \hfill
\caption{A portion of the horizontal plane $(r, \phi)$ cutting through the disc at $z/H_0 = 2$ comparing perturbation vertical vorticity, $\delta\omega_z$, for a 3D VSI run with and without Fargo.  Each run was started at $t/T_0 = 300.27$ and run for $0.5 T_0$ up to $t/T_0 = 300.77$. (a) With Fargo.  (b) Without Fargo.}
\label{fig:vsi3d}
\end{figure}

Figure~\ref{fig:vsi3d} compares the vertical vorticity ($\delta\omega_z$) perturbation (relative to the basic state) in the $z/H_0 = 2$ plane after a time of $0.5 T_0$, i.e., at $t/T_0 = 300.77$.  Here $T_0$ is the orbital period at the mid-radius of the computational domain.  The difference in the two solutions at this time is very small.  Due to the chaotic nature of the flow,  the error due to the different time steps, $\Delta t$, chosen for the two simulations grows with time and differences become more apparent.
The CFL number was chosen to be 1.5 and based on this,
the time step selected by the code for the non-Fargo run was $\Delta t/T_0 = 2.25 \times 10^{-4}$.  This choice was constrained by Keplerian advection.  For the Fargo run, the code selected $\Delta t/T_0 = 5.49 \times 10^{-4}$ for the same CFL number which represents a better than factor of two improvement.  In the run with Fargo, the choice of time step was constrained by the characteristic-wave speed and grid size in the radial direction.  The cpu time for the non-Fargo and Fargo runs was $1.50$ and $1.76$ seconds per step.  This represents a 17.3\% overhead for a more than factor of two gain in time step.  The Intel Broadwell processor was used for these runs.

We close with a brief description of the physics observed in the 3D VSI runs.
The Keplerian mean flow is counter-clockwise in Figure~\ref{fig:vsi3d} and the vertical vorticity perturbation consists of layers of cyclonic $\delta\omega_z$ (red bands) which induce across them, a positive jump in specific angular momentum, $j_\phi = u_\phi r$ as $r$ increases.    These layers are formed by the vertical transport of basic state angular momentum by the vertical jets that are the main feature of VSI.  This will be discussed in more detail in a forthcoming paper.  In between the cyclonic layers, one observes weaker and more diffuse anti-cyclonic $\delta\omega_z$ which reduces $j_\phi$ compared to the basic state.  In the inner portion of the disk, anti-cyclonic $\delta\omega_z$ structures take the form of smaller aspect ratio structures.

\subsection{Parallel scaling tests for 3D vertical shear instability runs}
\label{sec:scaling}

Following the suggestion of the referee, the efficiency $\eta$ of the code is defined as the ratio of the useful work to the resources consumed.  The useful work is the number $\Ngrid$ of grid points evolved while the resources consumed is the cpu time $\tcpu$ (per time step in secs.) times the number of cores, $\Ncores$.  We also include the ratio $\Delta t_\mathrm{actual} / \Delta t_\mathrm{Fargo}$ to account for the reduced time step in non-Fargo runs:
\be
   \eta \equiv \frac{\Ngrid/2^{17}}{\tcpu \Ncores} \frac{\Delta t_\mathrm{actual}}{\Delta t_\mathrm{Fargo}}, \eql{eta}
\ee
where we have normalized the number of grid points to $2^{17}$.
When there is perfect scaling, $\eta$ should be a constant as $\Ncores$ is increased.

\begin{figure}
%
\centerline{                 
\includegraphics[width=3.4in]{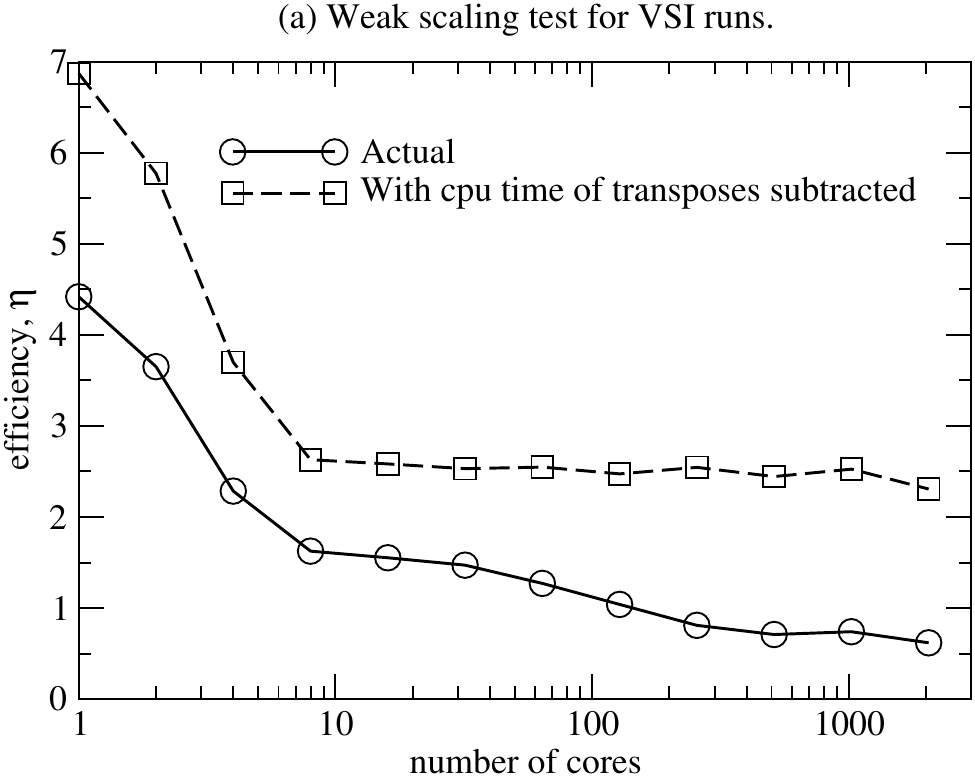} \hfill
\includegraphics[width=3.4in]{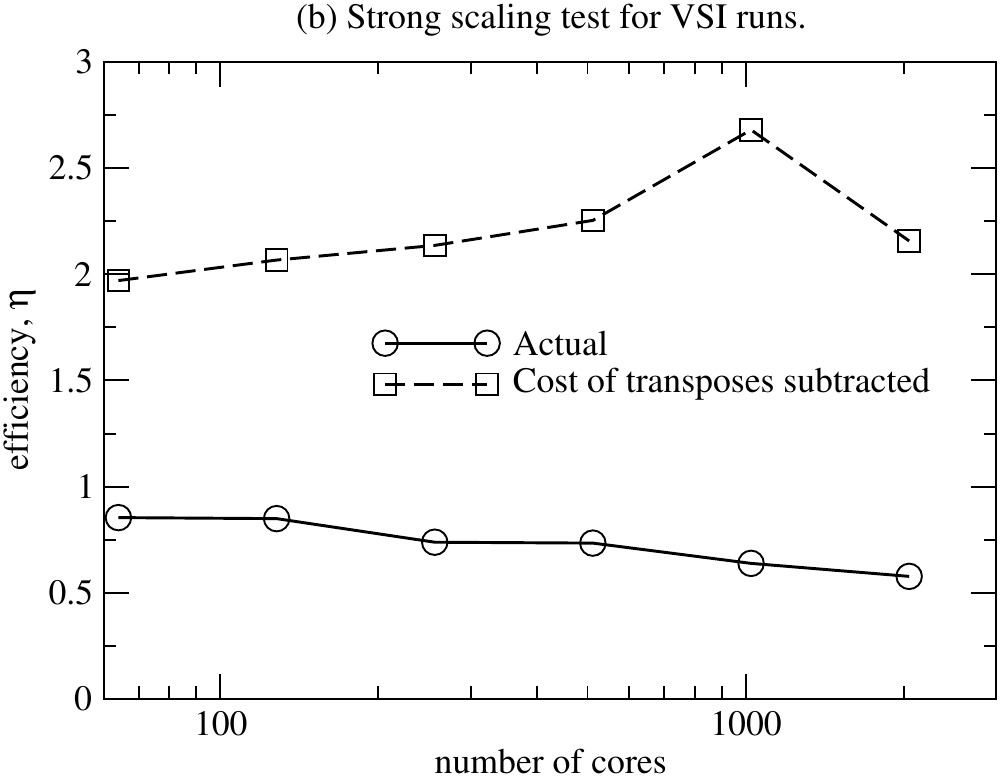}
                 }
\centerline{
\includegraphics[width=3.4in]{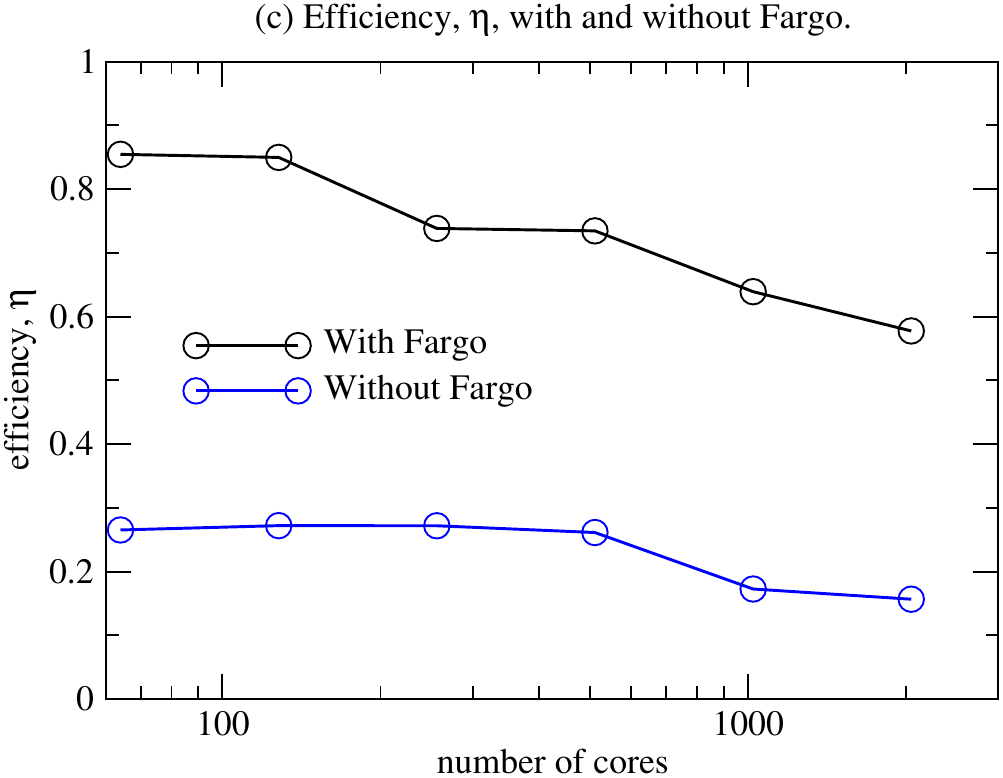}
                  }
\caption{Scaling tests for the 3D vertical shear instability (VSI) set-up performed using the Intel Haswell nodes.  The environment variable {\tt MPI\_IB\_RAILS} used by the InfiniBand network was set to 2 for all runs. (a) Efficiency, $\eta$, for the weak scaling test where the problem size per processor is fixed.  (b) Efficiency for the strong scaling test where the total problem size is fixed at $n_r \times n_z \times n_\phi = 512 \times 512 \times 1024$. (c) Efficiency, $\eta$, with Fargo deactivated.  The measure, $\eta$, accounts for the decrease in time step when Fargo is deactivated.}
\label{fig:scaling}
\end{figure}

All the tests in this sub-section were performed using Intel Haswell nodes which use the E5-2680V3 (Xeon) processor.
The variable {\tt MPI\_IB\_RAILS} used by the InfiniBand network was set to 2; this causes two InfiniBand (IB) fabrics to be used for communication and results in reduced cpu time.
Figures~\ref{fig:scaling}a and b plot $\eta$ for a 3D vertical shear instability (VSI) setup with Fargo activated; therefore the ratio $\Delta t_\mathrm{actual} / \Delta t_\mathrm{Fargo}$ in \eqp{eta} equals unity.

Figure~\ref{fig:scaling}a is for a weak scaling test in which the grid size per core is kept fixed while $\Ncores$ is increased.  In other words, both $\Ncores$ and the total grid size increase simultaneously.  In the present test, the number of grid points for the one core run is $64 \times 32 \times 64$ ($n_r \times n_z \times n_\phi$) and each direction is successively doubled in resolution as $\Ncores$ doubles.  Figures~\ref{fig:scaling}a plots two curves, one which used the actual cpu time (solid) and another (dashed) for which the time taken to perform transposes was subtracted out.  
Both curves show an initial rapid decrease in $\eta$ up to $\Ncores = 8$.
The main cause of this is likely the fact that each Xeon processor in a Haswell node has 12 cores which share a single memory and level 3 cache.  Contention for both resources increases as the number of cores increases from 1 to 12.
Therefore, we focus on the region $\Ncores \geq 12$.  At $\Ncores = 2048$, the efficiency has decreased to 38\% of its value at $\Ncores = 16$.  When the cpu time for transposes is subtracted out, the efficiency remains relatively flat.  This indicates that the loss in efficiency is due to communication intensive transposes.  Indeed, the fraction of time (not shown) taken for transposes increases from $0.38$ to $0.73$ as $\Ncores$ increases from 16 to 2048.

Figure~\ref{fig:scaling}b is for a strong scaling test in which the total problem size is fixed at $512 \times 512 \times 1024$ ($n_r \times n_z \times n_\phi$) and the number of cores is varied.  At $\Ncores = 2048$, the efficiency has decreased to 67\% of its value at $\Ncores = 64$.  The cpu fraction taken for transposes increases from 0.57 to 0.73 (not shown) in this range.  When the cpu time for transposes is subtracted out, the efficiency increases slowly.  This is explained as follows.  In the strong test, the total number of memory fetches is constant with increasing $\Ncores$, however,
the number of cache hits (when needed data is found to be already in cache) is likely to be statistically higher since the number of cache slots per fetch is higher with more cores.

Figure~\ref{fig:scaling}c compares the efficiency of the Fargo versus the non-Fargo scheme.  It shows that there is at least a factor of 3.7 advantage to using Fargo, i.e., the overhead of the Fargo method is more than compensated by an increase in time step.

In conclusion, the all-to-all communication of transposes leads to a significant loss in efficiency as the number of cores is increased.  To reduce this communication overhead, an effort is underway to use a parallel Pad\'e algorithm \citep{Kim_etal_2021}.

\section{Closing Remarks}\label{sec:close}

A code has been developed that uses a fourth-order Pad\'e scheme to simulate hydrodynamic turbulence in protoplanetary disks. Pad\'e schemes are non-dissipative and have high resolving power.  Thus, with the same grid resolution, they are better able to capture fine scale vortical features compared to the dissipative shock capturing schemes employed in most astrophysics codes.  They also have better resolving power then central finite difference schemes of the same order.

Suggested improvements are as follows.  (i) To eliminate communication intensive transposes, consider using parallel tridiagonal matrix algorithms \citep[][and the references therein]{Kim_etal_2021} which require much less communication.  \cite{Kim_etal_2021} demonstrate good scaling as the number of cores is increased.  (ii). For simulations that require long radial domains (more than $\approx 6$ scale heights) it would be better to use spherical rather than cylindrical coordinates.  An option for this could be provided.  (iii) The sixth-order tridiagonal Pad\'e scheme
\be
   \alpha f^\prime_{j-1} + f^\prime_j + \alpha f^\prime_{j+1} = \frac{a}{2h}\left(f_{j+1} - f_{j-1}\right)
   + \frac{b}{4h} \left(f_{j+2} - f_{j-2}\right), \eql{sixth_pade}
\ee
with $\alpha = 1/3$, $a = 14/9$, and $b= 1/9$ could be implemented. (iv) The capability to track Lagrangian solid particles could be provided.  (v) An intercomparison effort with other codes could be performed.

\vskip 0.5truecm
The code is available at \url{https://github.com/NASA-Planetary-Science/Pade-disk-code} and
\url{https://zenodo.org/records/11114378}.  The citation for Zenodo is \citet{Pade_Zenodo_2024}.

\begin{acknowledgments}

I am grateful to Alan Wray (NASA Ames) for providing several routines that have been used in {\sc Pad\'e}, including directional transposes, tridiagonal solvers, and the Pad\'e filter.  He also helped to explain features observed in the scaling tests.
I am grateful to Jeff Cuzzi (NASA Ames) for performing the internal review and for his encouragement, Debanjan Sengupta (New Mexico State Univ., Las Cruces) for performing the internal review and suggesting the weak scaling test; Orkan Umurhan (SETI Inst.) for his encouragement and advice on the vertical shear instability runs, Prof.~Ali Mani (Stanford Univ.) for advice on the implementation of artificial bulk viscosity, and Prof. Daniel Lecoanet (Northwestern Univ.) for providing the result of the Kelvin-Helmholtz test case run with {\sc Dedalus}. 
Finally, I thank the referee for many useful suggestions that improved the paper.
This work was partly supported by NASA's Internal Scientist Funding Model (ISFM) through the ``Origins'' group in NASA Ames' Space Science Division.

\end{acknowledgments}

\newpage
\appendix

\section{Molecular viscous force, viscous heating, and heat conduction}\label{sec:viscous_terms}

The code provides the option to add terms that represent molecular viscosity and heat conduction.
Similar terms arise in models of sub-grid turbulence.  For these we have coded the models due to \cite{Smagorinsky_1963} and \cite{Vreman_2004}.  However, since we have not tested them, this section describes the implementation for the molecular/laminar case only. If the Fargo option has been activated, Fargo chain-rule is applied wherever needed.

\subsection{Viscous force}

The viscous stress tensor is
\be
   T_{ij} = 2\mu S_{ij} + \left(\mu_\rmb - \frac{2}{3}\mu\right) S_{kk} \delta_{ij}, \eql{Tij}
\ee
where $\mu$ and $\mu_\rmb$ are the shear and bulk viscosities, respectively, and
\be
   S_{ij} = \frac{1}{2}\left(\p_j u_i + \p_i u_j\right)
\ee
is the strain rate tensor.

Equation \eqp{Tij} is implemented in {\tt subroutine laminar\_stress\_and\_heat\_flux}.
The components of the strain tensor (which is symmetric) are computed in
{\tt subroutine strain\_tensor} as follows
\ba
&&S_{zz} = \p_z u_z, \eqspace S_{\phi z} = \half\left(\p_z u_\phi + \ri \p_\phi u_z\right), \eqspace
S_{zr} = \half\left(\p_r u_z + \p_z u_r\right), \\
&&S_{rr} = \p_r u_r, \eqspace S_{r\phi} = \frac{1}{2}\left(\ri\p_\phi u_r + \p_r u_\phi - \frac{u_\phi}{r}\right),
\eqspace S_{\phi\phi} = \ri\p_\phi u_\phi + \frac{u_r}{r}.
\ea
The above expressions are from Aris (\citeyear{Aris_1989}, p.181) and agree with Batchelor(\citeyear{Batchelor_1967}, p. 602).

The viscous force is the divergence of the viscous stress tensor and given by Aris (\citeyear{Aris_1989}, p. 179) for orthogonal coordinates as follows (in his notation):
\be
F^\rmv_i = T(ij, j) = \frac{h_i}{h_1 h_2 h_3}\frac{\p}{\p x_j}\left[\frac{h_1 h_2 h_3}{h_i h_j} T(ij)\right] + \frac{h_i}{h_j h_k}
\left\{\substack{i \\j k}\right\} T(jk), \eqspace(\mathrm{no\ sum\ on\ }i). \eql{aris}
\ee
Here $(x_1, x_2, x_3) = (r, \phi, z)$ and corresponding scale factors are $h_1 = h_r = 1$, $h_2 = h_\phi = r$, and $h_3 = h_z = 1$.
The quantities in braces are Christoffel symbols and the only non-zero ones are
\be
\left\{\substack{2 \\1 2}\right\} = \left\{\substack{2 \\2 1}\right\} = r \mathrm{\ and\ } \left\{\substack{1 \\2 2}\right\} = -r. 
\ee

The symbolic algebra package maxima was used to verify the correctness of Aris' expression \eqp{aris} by writing Cartesian velocities in terms of cylindrical quantities (velocities and coordinates) and computing Cartesian viscous forces in terms of cylindrical quantities using the chain rule throughout.  These can be be rotated to obtain the forces in cylindrical coordinates in terms of cylindrical quantities and compared with \eqp{aris}.  The relevant maxima script can be found in {\tt check\_Aris.mac} in the
{\tt Symbolic\_algebra} sub-directory.

The final expressions for the viscous force area:
\ba
F^\rmv_z      &=& \ri\left[\p_r(r T_{zr})+ \p_\phi(T_{z\phi} )+ \p_z(r T_{zz})\right], \\
F^\rmv_r      &=& \ri\left[\p_r(r T_{rr}) + \p_\phi(T_{r\phi})  + \p_z(r T_{rz})\right] - \ri T_{\phi\phi}, \\
F^\rmv_\phi &=&             \p_r(T_{\phi r}) + \ri \p_\phi(T_{\phi\phi}) + \p_z(T_{\phi z}) + \frac{2}{r} T_{\phi r}.
\ea
These agree with the expressions given on p. 739B of \cite{BSL}.

\subsection{Viscous heating}

The equation for total energy $e = e_\mathrm{int} + \rho u_i u_i$ (per unit volume) has a term for the work done by shear stresses
(Liepmann \& Roshko, \citeyear{Liepmann_and_Roshko_2001}, p. 335),
\be
   W^\mathrm{shear} = \frac{\p}{\p x_j} (T_{ij} u_i). 
\ee
For the internal energy equation which we solve, we must subtract the kinetic energy dissipation:
\be
   D = u_i \frac{\p}{\p x_j} T_{ij}.
\ee
This gives the viscous heating term for the internal energy
\be
   Q^\rmv = W^\mathrm{shear} - D = T_{ij} \frac{\p u_i}{\p x_j}. \eql{Qv}
\ee
Now $T_{ij}$ is symmetric so only the symmetric part of $\p_j u_i$ survives.  Then substituting the constitutive equation \eqp{Tij} for $T_{ij}$ into \eqp{Qv} one obtains
\be
   Q^\rmv = 2\mu S_{ij} S_{ij} + \left(\mu_\rmb - \frac{2}{3}\mu\right) S_{kk}^2. \eql{Qv_final}
\ee

\subsection{Heat conduction}

The flux of internal energy due to molecular conductivity is given by Fourier's law:
\be
   \qveccond = - \frac{k}{c_v} \nabla(c_v T),
\ee
where we have multiplied and divided by $c_v$ inside and outside the gradient, respectively.
This assumes that $c_v$ is constant, i.e., that we have a calorically perfect gas.  Now $c_v T$ is simply $\eint /\rho$, and using the definition of the Prandtl number $\mathrm{Pr} \equiv \mu c_p / k$ we get
\be
\frac{k}{c_v} = \frac{\mu\gamma}{\mathrm{Pr}}.
\ee

\section{Discrete conservation} \label{sec:cons}

\subsection{Theory}

Here we describe how to choose boundary schemes to ensure that the overall scheme possesses a discrete conservation property.  We follow \cite{Lele_1992} and \cite{Brady_and_Livescu_2019} and offer two clarifications:  (1) There is a distinction between provisional and final weights; the weights given in \S4.2 of \cite{Lele_1992} are provisional.  (2) The weights cannot be specified \textit{a priori} and must be determined as part of the solution.  Specifically, one needs to verify that the final weights provide a reasonable discrete conservation law.  In general, the final weights will not correspond exactly to a quadrature rule. 

For a system of partial differential equations in more than one dimension, we compute derivatives of fluxes along each direction separately.  Hence, it is sufficient to consider the one-dimensional partial differential equation
\be
\frac{\p u}{\p t} + \frac{\p f}{\p x}, \hskip 0.5truecm x \in [0, L]. \eql{pde}
\ee
Upon integration over the domain, \eqp{pde} gives the conservation law:
\be
\frac{d}{dt}\int_0^L u(x, t)\, dx = f(0) - f(L). \eql{cons_law}
\ee
Pad\'e differencing applied to \eqp{pde} should possess a discrete analog of \eqp{cons_law}; in its absence, a long time solution can drift and fail to achieve statistical stationarity.  
Pad\'e difference schemes have the form
\be
   A \bffp = B \bff.
\ee
where $A$ and $B$ are banded matrices
and henceforth, lower case bold letters will be used to denote column vectors.

We now state a result that was stated by \cite{Lele_1992} without a proof, which was later provided by
\cite{Brady_and_Livescu_2019}.

\begin{proposition}
To obtain a discrete analog of \eqp{cons_law}, columns 2 through $N-1$ of matrix $B$ must have a weighted sum of zero, i.e.,
\be
  \bfw^T \bfb_i = 0 \hskip 0.25truecm \mathrm{for\ } i \in [2, N-1], \eql{governing_for_weights}
\ee
where $\bfw$ is a column vector of weights and $\bfb_i$ is the $i$th column of matrix $B$.
The weights $\bfw$ are provisional; final weights will be given below.
\end{proposition}
\begin{remark}
Not all the weights, $\bfw$, can be specified \textit{a priori} but must be obtained as part of the process of satisfying \eqp{governing_for_weights}.
\end{remark}

\begin{proof}

We assume that grid points $x_i, i = 1, \ldots, N$ have been laid out according to a smooth analytic mapping $x(\xi)$ such that $\xi_i = i$, i.e., $\xi$ is a smooth grid index variable.  In implementations, it is convenient to take derivatives with respect to $\xi$ and then use the chain rule:
\be
\left(\frac{\p f}{\p x}\right)_i = \left(\frac{\p f}{\p\xi}\right)_i \left(\frac{d\xi}{dx}\right)_i.
\ee
For simplicity we use the notation
\be
   f_i^\prime(\xi) \equiv \left(\frac{\p f}{\p\xi}\right)_i, \hskip 0.5truecm h_i^{-1} \equiv \left(\frac{d\xi}{dx}\right)_i.
\ee
Then the spatially discretized version of \eqp{cons_law} is
\be
  \frac{du_i}{dt} + \frac{1}{h_i}\sum_{j,k} \left[A^{-1}\right]_{ij} B_{jk}f_k = 0. \eql{disc_cons_law}
\ee
Defining the vector $\bfU = \left[h_1 u_1, h_2 u_2, \ldots, h_N u_N\right]^T$, we can write \eqp{disc_cons_law} as
\be
A \frac{d\bfU}{dt} + B\bff = 0. \eql{sd}
\ee
A weighted sum is applied to \eqp{sd} to mimic the integration in \eqp{cons_law}:
\be
\frac{d}{dt} \bfw^T A \bfU = - \bfw^T B \bff. \eql{sd_quad}
\ee
Following \cite{Brady_and_Livescu_2019}, let $B = [\bfb_1, \bfb_2, \ldots, \bfb_N]$ where $\bfb_i$ denotes the $i$th column vector of the matrix $B$. Now
\ba
\bfw^T B &=& \bfw^T [\bfb_1, \bfb_2, \ldots, \bfb_N], \\
               &=& [\bfw^T\bfb_1,\bfw^T\bfb_2, \ldots, \bfw^T\bfb_N], \\
\ea
so that
\be
   \bfw^T B \bff = \sum_{i=1}^N \bfw^T \bfb_i f_i.
\ee
Hence \eqp{sd_quad} can be written
\be
   \frac{d}{dt} \bfw^T A \bfU = -\bfw^T \bfb_1 f_1 - \bfw^T\bfb_N f_N - \sum_{i=2}^{N-1} \bfw^T \bfb_i f_i. \eql{sd3}
\ee
In order to arrive at a discrete conservation law analogous to \eqp{cons_law}, let us try imposing
\be
\sum_{i=2}^{N-1} \bfw^T \bfb_i f_i = 0.  \eql{this}
\ee
For \eqp{this} to be true for arbitrary $f_i$ we must have that
\be
  \bfw^T \bfb_i = 0 \hskip 0.25truecm \mathrm{for\ } i \in [2, N-1]. \eql{governing}
\ee
In other words we want columns 2 through $N-1$ of the matrix $B$ to have a weighted sum of zero.  

We now show, finally, that the trial condition \eqp{governing} does indeed lead to a discrete conservation law.  Equation \eqp{governing} gives the set of conditions one uses to solve for the provisional weights.
Then \eqp{sd3} becomes
\be
   \frac{d}{dt} \bfwt^T \bfU = -\bfw^T \bfb_1 f_1 - \bfw^T\bfb_N f_N, \eql{sd4}
\ee
where we have defined a new set of weights
\be
   \bfwt^T \equiv \bfw^T A.
\ee
Now $\bfw^T \bfb_1$ is a number and if we use the same boundary schemes and quadrature weights at the left and right boundaries then $-\bfw^T \bfb_1 = + \bfw^T \bfb_N$ and we can divide \eqp{sd4} by it.  Finally, putting back $U_i = u_i h_i$ \eqp{sd4} we get the desired discrete conservation law
\be
   \frac{d}{dt} \sum_{i=1}^N \wh_j h_j u_j = f_1 - f_N, \eql{discrete_analog_final},
\ee
where the final weights are
\be
   \widehat{\bfw} = - \frac{\bfw^T A}{\bfw^T \bfb_1}. \eql{final}
\ee
Note that this expression is homogeneous in the provisional weights $\bfw$.  Hence we may scale the provisional weights as we wish when we solve equations \eqp{governing} to determine them.  In particular, it it is convenient to choose them to be unity in the interior of the domain.
\end{proof}

\subsection{Application to the present differentiation scheme}

We consider matrix entries near the left end of the domain; the weights near the right end are obtained by symmetry.
Referring back to \S\ref{sec:nbc}, the first five columns of the right-hand-side matrix of the present Pad\'e scheme are 
\be
B = \left(\begin{matrix}
a_1    & b_1   & c_1   &         &\\
-\ahat & 0       & \ahat &         &\\
          & -\ahat & 0      & \ahat & \\
          &            & -\ahat & 0 & \ahat\\
          &            &           & -\ahat & 0\\
          &            &           &           & -\ahat\\
\end{matrix}\right),
\ee
where we have defined $\ahat = a/2$.
The condition that the weighted sum of the fifth column equals zero implies that
$w_4 = w_6$ which we set equal to unity.  The same holds true for the rest of the interior weights.
The condition on the fourth column implies that $w_3 = w_5$ which we also set equal to unity.
Using the fact that $w_3 = 1$, the second column gives the condition
\be
  w_1 b_1 - \ahat = 0,
\ee
or 
\be
   w_1 = \ahat/b_1 = (3/4) / 2 = 3/8.
\ee
The third column gives
\be
   w_1 c_1 + w_2 \ahat - w_4 \ahat = 0.
\ee
Using known information, we get
\be
   w_2 = 1 - c_1/b_1 = 3/4.
\ee
We now have all the provisional weights.

The final weights are obtained from \eqp{final} where the first four columns of the LHS matrix for the present scheme is
\be
   A = \left(
   \begin{matrix}
   1 & \alpha_1 & 0 &\\
   \alpha & 1 & \alpha& \\
   & \alpha & 1 &\alpha\\
   &&\alpha&1&\\\
   &&&\alpha
   \end{matrix}\right)
\ee
Hence \eqp{final} gives the final weights as
\begin{align}
   \widehat{w}_1 &= -\frac{w_1 + w_2\alpha}{w_1 a_1 - w_2 \ahat} = \frac{3}{8}\\
   \widehat{w}_2 & = -\frac{w_1\alpha_1 + w_2 + w_3\alpha}{w_1 a_1 - w_2\ahat} = \frac{7}{6}\\
   \widehat{w}_3 & = -\frac{w_2\alpha + w_3 + w_4\alpha}{w_1 a_1 - w_2\ahat} = \frac{23}{24}\\
   \widehat{w}_j & = - \frac{w_{j-1}\alpha + w_j + w_{j+1}\alpha}{w_1 a_1 - w_2\ahat} = 1, \hskip 0.5truecm j \in [4, N-3].
\end{align}
It is interesting that $\what_1 + \what_2 + \what_3 = 5/2$ which would be the case for trapezoidal rule and indeed the three weights approximate the weights $1/2$, $1$, and $1$ for trapezoidal rule.

\bibliography{code.bbl}

\end{document}